\documentclass{aa}  
\usepackage{graphicx}
\usepackage{multirow}
\usepackage{upgreek}
\usepackage{natbib}

\usepackage{txfonts}
\begin{document}

\title{VLT/SPHERE survey for exoplanets around young early-type stars, including systems with multi-belt architectures}
\titlerunning{Searching for Exoplanets around Young, Early-type Stars}

\author{
M. Lombart\inst{1,2},
G. Chauvin\inst{1,3},
P. Rojo\inst{4},
E. Lagadec\inst{5}
P. Delorme\inst{3},
H. Beust\inst{3},
M. Bonnefoy\inst{3},
R. Galicher\inst{6},
R. Gratton\inst{7},
D. Mesa\inst{7},
M. Bonavita\inst{8},
F. Allard\inst{9},
A. Bayo\inst{10,11},
A. Boccaletti\inst{6},
S. Desidera\inst{7},
J. Girard\inst{12},
J.S. Jenkins\inst{4},
H. Klahr\inst{15},
G. Laibe\inst{2}, 
A.-M. Lagrange\inst{3},
C. Lazzoni\inst{7},
G-D.~Marleau\inst{13,14,15},
D. Minniti\inst{16},
and C. Mordasini\inst{14}}
\authorrunning{Lombart et al.}

\institute{
$^{1}$ Unidad Mixta Internacional Franco-Chilena de Astronom\'{i}a, CNRS/INSU UMI 3386 and Departamento de Astronom\'{i}a, Universidad de Chile, Casilla 36-D, Santiago, Chile\\
$^{2}$ Univ Lyon, Univ Lyon1, Ens de Lyon, CNRS, Centre de Recherche Astrophysique de Lyon UMR5574, F-69230, Saint-Genis,-Laval, France \\
$^{3}$ Univ. Grenoble Alpes, CNRS, IPAG, F-38000 Grenoble, France. \\
$^{4}$ Departamento de Astronom\'ia, Universidad de Chile, Camino El Observatorio 1515, Las Condes, Santiago, Chile
$^{5}$  Universit\'e Cote d'Azur, OCA, CNRS, Lagrange, France\\
$^{6}$ LESIA, Observatoire de Paris, PSL Research University, CNRS, Sorbonne Universit\'es, UPMC Univ. Paris 06, Univ. Paris Diderot, Sorbonne Paris Cit\'e, 5 place Jules Janssen, 92195 Meudon, France\\
$^{7}$ INAF - Osservatorio Astronomico di Padova, Vicolo dell’ Osservatorio 5, 35122, Padova, Italy\\
$^{8}$ SUPA, Institute for Astronomy, The University of Edinburgh, Royal Observatory, Blackford Hill, Edinburgh, EH9 3HJ, UK\\
$^{9}$ CRAL, UMR 5574, CNRS, Universit\'e de Lyon, Ecole Normale Suprieure de Lyon, 46 Alle d'Italie, F-69364 Lyon Cedex 07, France\\
$^{10}$ Instituto de F\'isica y Astronom\'ia, Facultad de Ciencias, Universidad de Valpara\'iso, Av. Gran Breta\~na 1111, Valpara\'iso, Chile\\
$^{11}$ N\'ucleo Milenio Formaci\'on Planetaria - NPF, Universidad de Valpara\'iso \\
$^{12}$ Space Telescope Science Institute, 3700 San Martin Dr. Baltimore, MD 21218, USA\\
$^{13}$ Institut f\"ur Astronomie und Astrophysik, Universit\"at T\"ubingen,Auf der Morgenstelle 10,D-72076 T\"ubingen, Germany\\ \label{Tue} 
$^{14}$ Physikalisches Institut,Universit\"{a}t Bern,Gesellschaftsstr.~6,CH-3012 Bern, Switzerland \label{Bern} \\
$^{15}$ Max-Planck-Institut f\"ur Astronomie,K\"onigstuhl 17,D-69117 Heidelberg, Germany \label{MPIA}\\
$^{16}$ Depto. de Ciencias Fisicas, Facultad de Ciencias Exactas, Universidad Andres Bello, Av. Fernandez Concha 700, Las Condes, Santiago, Chile\\
\email{maxime.lombart@ens-lyon.fr}
}

\abstract
   {Dusty debris disks around pre- and main-sequence stars are potential signposts for the existence of planetesimals and exoplanets. Giant planet formation is therefore expected to play a key role in the evolution of the disk. This is indirectly confirmed by extant submillimeter near-infrared images of young protoplanetary and cool dusty debris disks around main-sequence stars that usually show substantial spatial structures. With two decades of direct imaging of exoplanets already studied, it is striking to note that a majority of recent discoveries of imaged giant planets have been obtained around young early-type stars hosting a circumstellar disk.}
   {Our aim was to create a direct imaging program designed to maximize our chances of giant planet discovery and target 22 young early-type stars. About half of them show indications of multi-belt architectures. }
   {Using the IRDIS dual-band imager and the IFS integral field spectrograph of SPHERE to acquire high-constrast coronagraphic differential near-infrared images, we conducted a systematic search in the close environment of these young, dusty, and early-type stars. We used  a combination of angular and spectral differential imaging to reach the best detection performances down to the planetary mass regime.}
   {We confirm that companions detected around HIP 34276, HIP 101800, and HIP 117452 are stationary background sources and binary companions. The companion candidates around HIP 8832, HIP 16095, and HIP 95619 are determined as background contaminations. Regarding the stars for which we infer the presence of debris belts, a theoretical minimum mass for planets required to clear the debris gaps can be calculated. The dynamical mass limit is at least $0.1 M_J$ and can exceed $1M_J$. Direct imaging data is typically sensitive to planets down to $\sim 3.6M_J$ at 1”, and $1.7M_J$ in the best case. These two limits tightly constrain the possible planetary systems present around each target. These systems will be probably detectable with the next generation of planet imagers.}
   {}

   \keywords{Techniques: Imaging and spectroscopy - Planets and Satellites: detection, fundamental parameters, atmospheres}

\maketitle
%

\section{Introduction}

How giant planets form and evolve is one of the biggest challenges of modern astronomy and remains a subject of heated debate. This major goal is directly connected to the ultimate search for life over the horizon 2030 to 2040, although several astrophysical (formation, evolution, dynamics, structure, and atmosphere), biological (bio-markers), and technical (new technologies developed for next generation of instrumentation) steps must be carried out in that perspective. Understanding how giant planets are formed and structured, how they evolve and interact, is critical as they completely shape the planetary system architectures and therefore the possibility of forming telluric planets capable of hosting life. More than two decades ago, the only planets we knew were the ones of our Solar System. With the manna of exoplanet discoveries since the 51~Peg discovery \citep{Mayor1995}, the diversities of systems found (hot Jupiters, irradiated and evaporating planets, misaligned planets with stellar spin, planets in binaries, telluric planets in habitable zones, discovery of Mars-sized planets...), the theories of planetary formation have drastically evolved to digest these observing constraints. However, we are still missing the full picture, and some key fundamental questions still lack answers. For example: i/ the physical processes at play to pass the km-size barrier to form planetary cores, ii/ the physics of accretion to form planetary atmospheres, iii/ the formation mechanisms to explain the existence of giant planets at wide orbits, iv/ the physical properties of young Jupiters, v/ the impact of planet-planet and planet-disk interaction in the final planetary system architecture, or vi/ the influence of the stellar mass and stellar environment in the planetary formation processes. Neither core accretion plus gas capture (CA; \citealt{Pollack1996}) nor disk fragmentation driven by gravitational instabilities (GI; \citealt{Cameron1978}) can globally explain all current observables from planet hunting techniques. Alternative mechanisms are then proposed, such as pebbles accretion to enable core accretion to operate at wide orbits \citep{Lambrechts2012}, inward/outward migration or planet-planet \citep{Crida2009,Bromley2014} or simply the possibility to have several mechanisms forming giant planets \citep{Boley2009}. In this context, each individual discovery of a giant planet and young planetary system using direct imaging is rich in terms of scientific exploitation and characterization, as these systems offer the possibility of i/ directly probing the presence of planets in their birth environments, ii/ enabling the orbital, physical, and spectral characterization of young massive Jupiters, iii/ characterizing the population of giant planets at all separations in synergy with complementary techniques such as astrometry (\textit{GAIA}) and radial velocity adapted to filter stellar activity.\\
  
Dusty debris disks around pre- and main-sequence stars are possible signposts for the existence of planetesimals and exoplanets \citep{Matthews2014}. Numerous T Tauri and Herbig stars indicate that the characteristic timescale for the dispersal of a surrounding dusty, gaseous disk is a few million years \citep{Kennedy2008b}. Giant planet formation is therefore expected to play a key role in the evolution of disk. This is indirectly confirmed by extant submillimeter and near-infrared images of cool dusty debris disks around main-sequence stars usually showing substantial spatial structure (e.g., $\epsilon$ Eri, Vega, Fomalhaut, $\beta$ Pic; see \citealt{Schneider2014}). It is striking to note that a majority of recent discoveries of imaged giant planets have been obtained around young, dusty, early-type stars. It includes the breakthrough discoveries of Fomalhaut b (3~$M_{\rm{Jup}}$ at 110~AU, A4V star; \citealt{Kalas2008}), HR\,8799 bcde (5-10~$M_{\rm{Jup}}$ at 10-64~au, F0V star; \citealt{Marois2010}), $\beta$\,Pictoris\,b (8-13~$M_{\rm{Jup}}$ at 9~au, A5V star; \citealt{Lagrange2010}), HD\,95086\,b (3-5~$M_{\rm{Jup}}$ at 56~au, A8V star; \citealt{Rameau2013}), and more recently 51\,Eri\,b (2~$M_{\rm{Jup}}$ at 14~au, F0V star; \citealt{Macintosh2015}). The presence of dust and the spatial substructure (ring, gap, warp, and other asymmetries) are possible indirect indicators of the presence of giant planets \citep{Mouillet1997,Dipierro2015,Pinte2020}. Direct imaging is here a unique and viable technique to complete our view of planetary system characteristics at wide orbits ($\ge5$~au). This technique enables us to directly study the planet-disk connection to constrain the planet's and disk's physical properties, evolution, and formation. In the case of $\beta$\,Pictoris, \cite{Lagrange2012} confirmed that $\beta$\,Pic\,b was actually responsible for the disk inner warp geometry, perturbing the planetesimals field and shaping the warp up to 40-60~au. The stars HD\,95086 and HR\,8799 share a common two-component architecture consisting of a warm inner belt ($\le5~$au) and a cold outer disk ($100-200~$au) (see \citealt{Su2015}). \cite{Kennedy2014} actually showed that the spectral energy distributions of both systems are consistent with two-temperature components compatible with dust emission arising from two distinct radial locations.  Such an architecture would be analogous to the outer Solar System’s configuration of asteroid and Kuiper belts separated by giant planets. Therefore, following the strategy of our NaCo DUSTIES (Dusty, yoUng, and early-type STar Imaging for ExoplanetS) survey \citep{Rameau2013} that led to the discovery of HD\,95086\,b, we initiated a searching for giant planets with SPHERE at VLT around an newly identified sample of young early-type stars with indication for some cases of multi-belt architecture to maximize the chances of discoveries. The sample, the observations, and the data reduction and analysis are presented in Sections \ref{sec:target_prop}, \ref{sec:observations} and \ref{sec:data_reduc_analysis}, respectively. The results are reported in Section \ref{sec:cc_detection} and discussed in Section \ref{sec:detection_limits}. 
  
\section{Target Properties}
\label{sec:target_prop}

\begin{figure}
\centering
\includegraphics[width=0.45\textwidth]{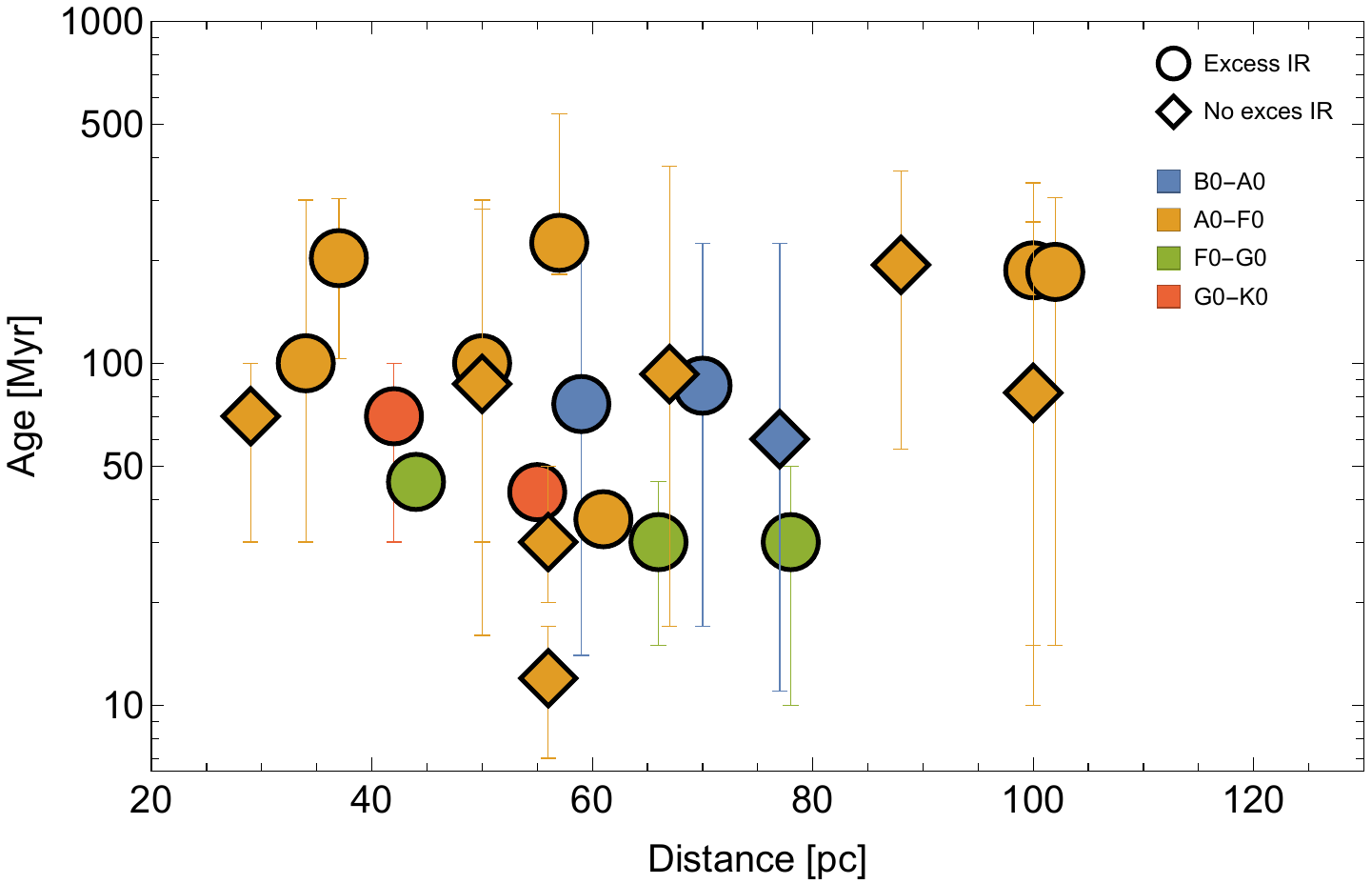}
\caption{Diagram of target properties taking into account age with error bars, distance, spectral type, and excess in infrared.}
\label{targets_prop}
\end{figure}

The target selection of the survey was obtained
from a large sample of young, nearby early-type stars  according to the following criteria: declination ($\delta \leq 25^{o}$), age ($\leq 100$ Myr), distance ($\leq 100$\,pc), and R-band brightness ($\leq 9.5$) to favor good adaptive optics performances. Age selection criteria were applied based on different youth diagnostics (kinematics, isochrones, Lithium, H$_\alpha$ emission, X-ray activity, stellar rotation, and chromospheric activity). We also used, as selection criteria, the presence of significant $60-70\,\mu$m excess from the \textit{IRAS} and \textit{Spitzer} missions in the spectral energy distributions \citep{Zuckerman1995,Zuckerman2001,Rhee2007,Zuckerman2004,Zuckerman2004b,Zuckerman2011,Zuckerman2013,David2015,Moor2016} or the existence of multi-belt component analysis from \cite{Kennedy2014}. A final total of 30 late-B-, A-, and early-F-type young stars, observable from the southern hemisphere, were then kept, 22 of which were observed between October 2016 and August 2019. Their stellar properties are reported in Table\,1. The age, distance, spectral type, and IR excess properties are shown in Figure~\ref{targets_prop}.

\begin{figure*}[t]
\centering

\includegraphics[width=0.3\textwidth]{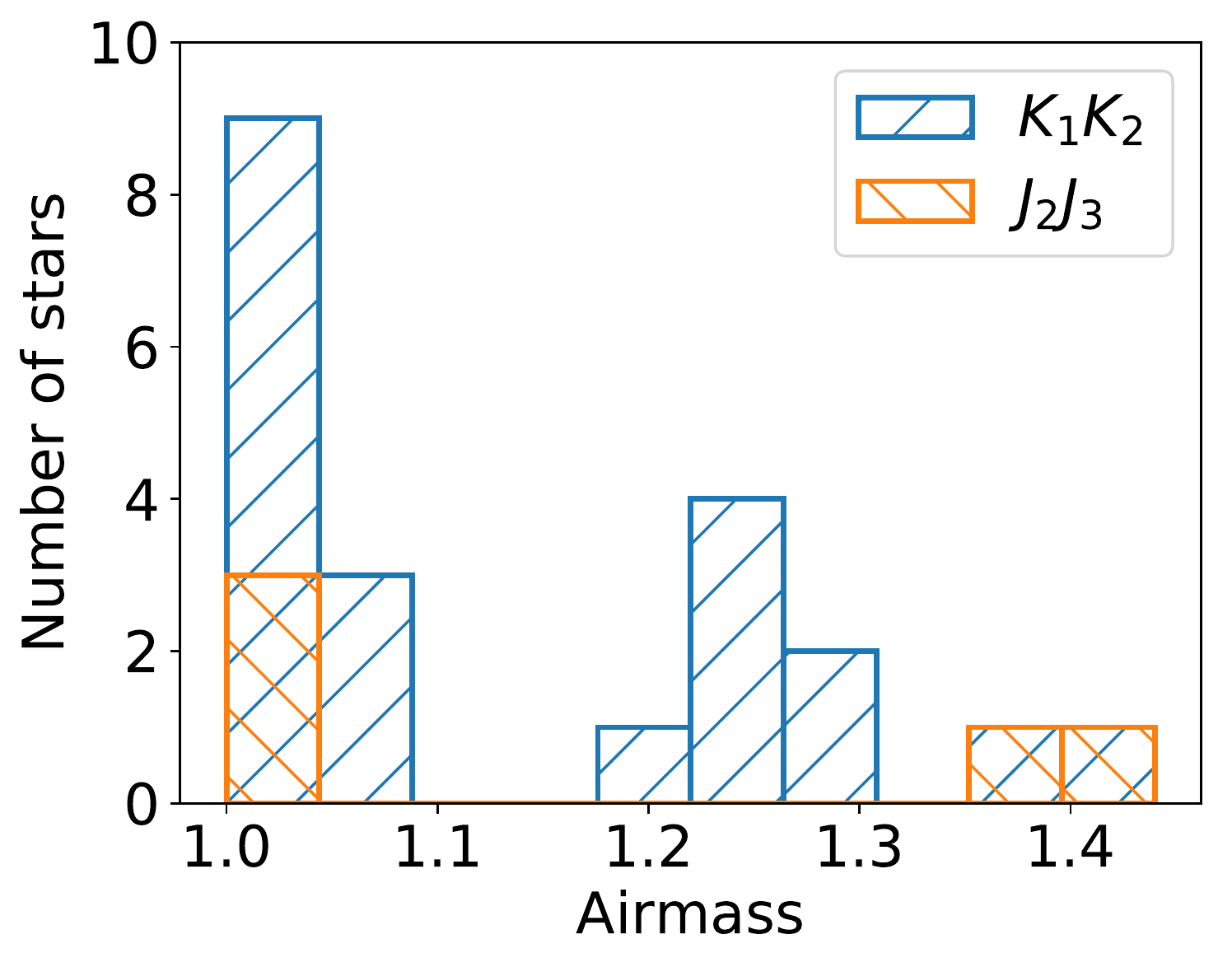}
\includegraphics[width=0.3\textwidth]{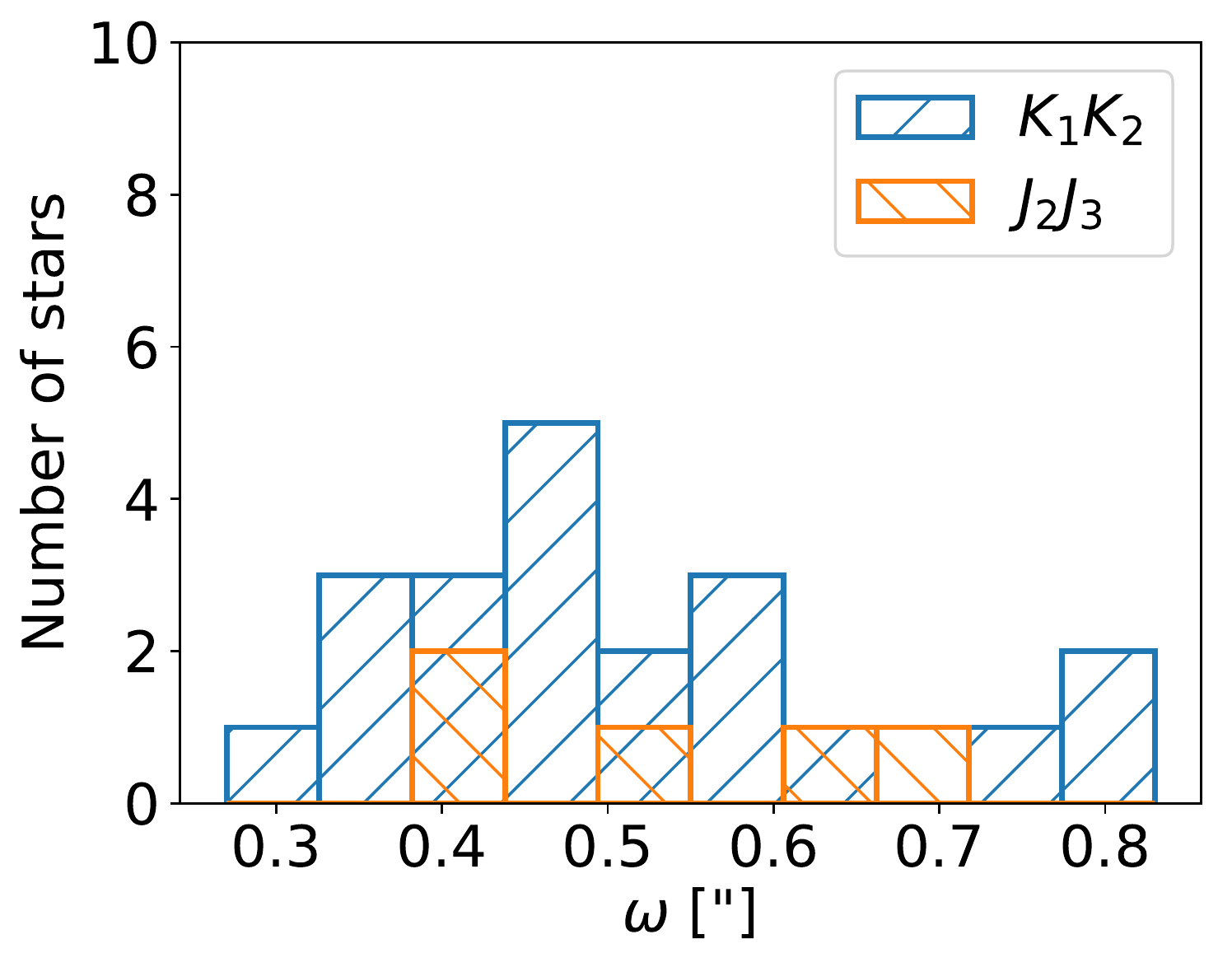}
\includegraphics[width=0.3\textwidth]{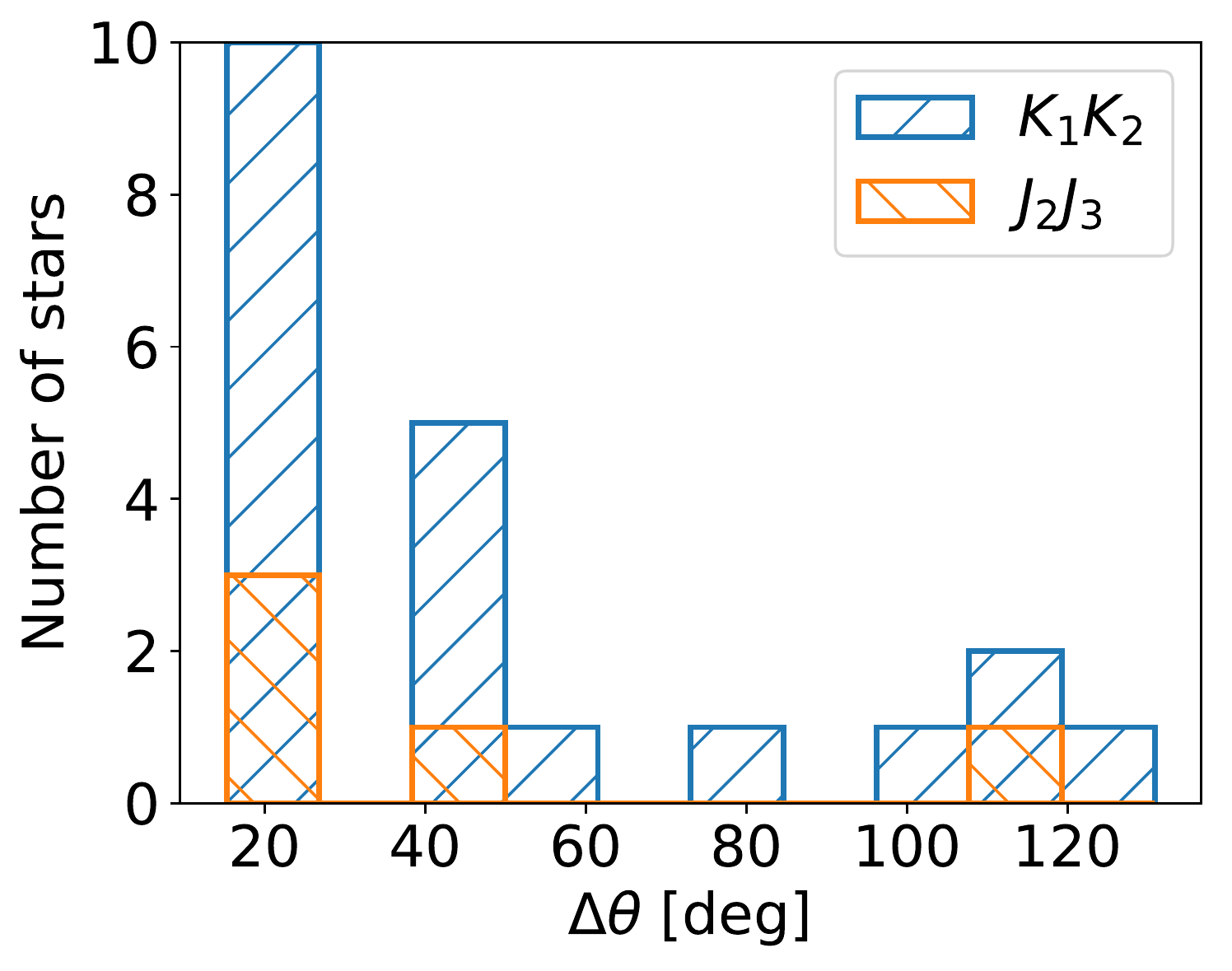}
\includegraphics[width=0.3\textwidth]{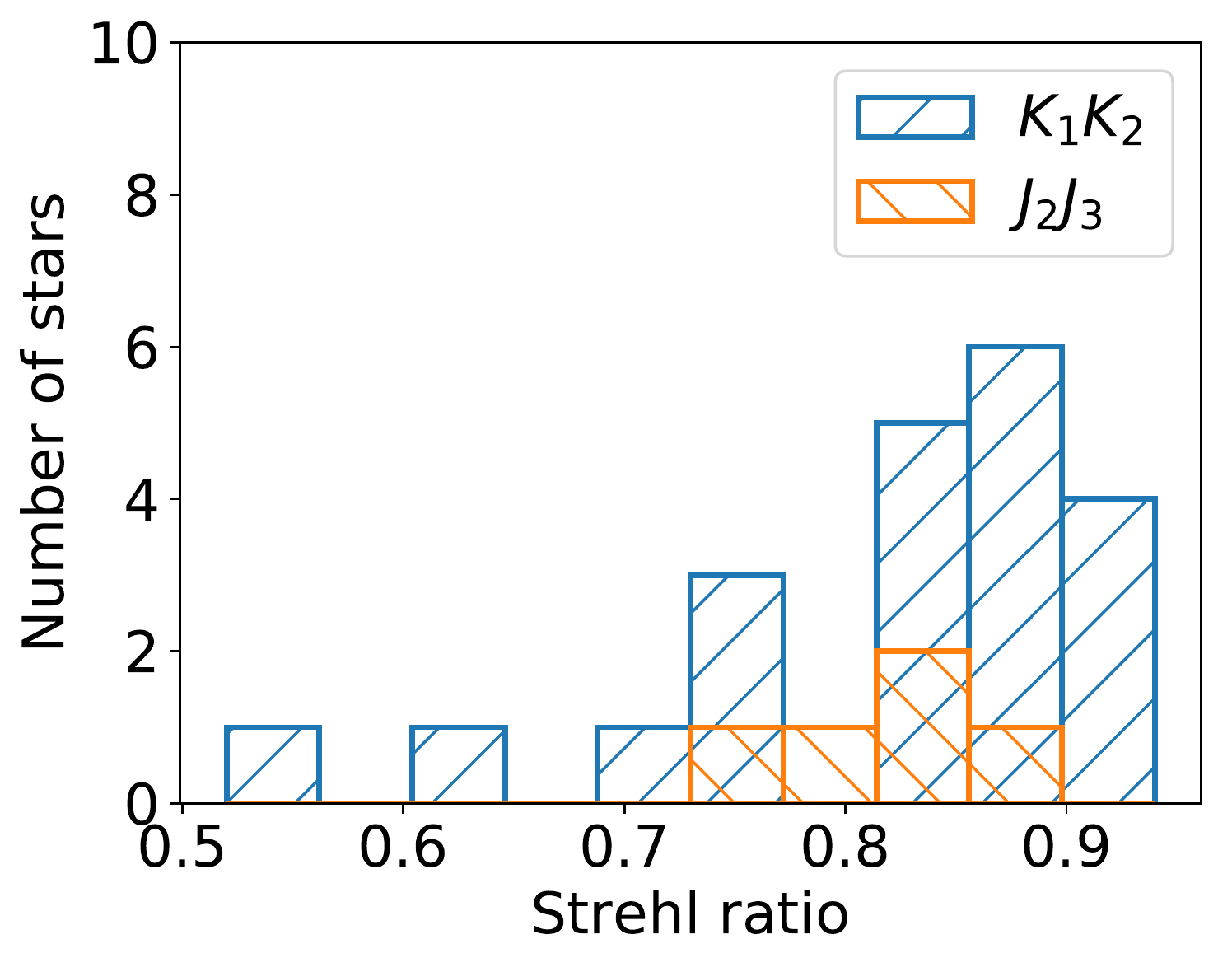}
\includegraphics[width=0.3\textwidth]{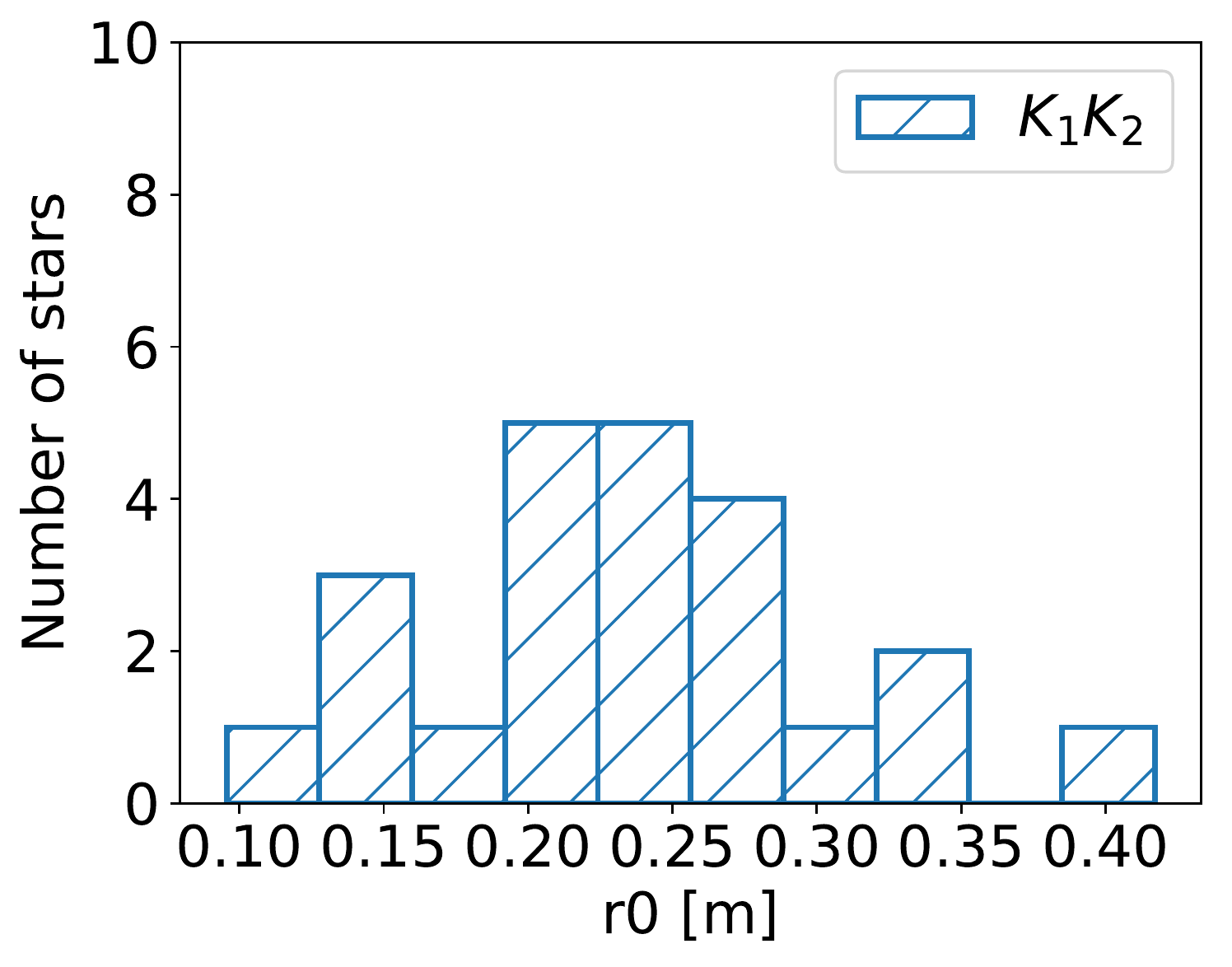}

\caption{Distribution of the SAXO real-time parameters, averaged over each observing sequence, for the complete survey: airmass, DIMM seeing ($\omega$), parallactic angle variation ($\Delta \theta$), the Strehl ratio at 1.6\,$\mu$m, and the Fried parameter of the atmosphere ($r_0$). }
\label{sparta}
\end{figure*}

\defcitealias{David2015}{D15}
\defcitealias{Zuckerman1995}{Z95}
\defcitealias{Zuckerman2001}{Z01}
\defcitealias{Zuckerman2004}{Z04}
\defcitealias{Zuckerman2004b}{Z04b}
\defcitealias{Zuckerman2011}{Z11}
\defcitealias{Zuckerman2012}{Z12}
\defcitealias{Zuckerman2013}{Z13}
\defcitealias{Rhee2007}{R07}
\defcitealias{Kennedy2014}{K14}
\defcitealias{Moor2016}{M16}
\defcitealias{Bell2015}{B15}
\defcitealias{Nielsen2019}{N19}
\defcitealias{Galicher2016}{G16}
\defcitealias{Meshkat2017}{M17}
\defcitealias{Vigan2017}{V17}

\begin{table*}

\caption{Description and properties of the sample. The Exc. column indicates the presence of an IR excess. The symbol "/" means no IR excess, and "Y" means with IR excess. References: \citepalias{Bell2015} \cite{Bell2015}; \citepalias{David2015} \cite{David2015}; \citepalias{Galicher2016} \cite{Galicher2016};  \citepalias{Kennedy2014} \cite{Kennedy2014}; \citepalias{Moor2016} \cite{Moor2016}; \citepalias{Meshkat2017} \cite{Meshkat2017}; \citepalias{Rhee2007} \cite{Rhee2007}; \citepalias{Vigan2017} \cite{Vigan2017}; \citepalias{Zuckerman1995} \cite{Zuckerman1995}; \citepalias{Zuckerman2001} \cite{Zuckerman2001}; \citepalias{Zuckerman2004} \cite{Zuckerman2004};\citepalias{Zuckerman2004b} \cite{Zuckerman2004b}; \citepalias{Zuckerman2011} \cite{Zuckerman2011}; \citepalias{Zuckerman2012} \cite{Zuckerman2012}; \citepalias{Zuckerman2013} \cite{Zuckerman2013}. }

\centering

\begin{tabular}{lllllllllll}

\hline 
Target & RA(2000) & DEC(2000) & $\mu_{\alpha}$ & $\mu_{\delta}.\cos(\delta)$ & H & SpT & Dist. & Age & Exc. & References \\

&&& (mas/yr) & (mas/yr) & (mag) && (pc) & (Myr) &&\\
\hline
\noalign{\vskip 1mm}
HIP3277 & 00 41 46.3 & -56 30 04.73  & 90.79 & 57.19 & 5.6 & A3V  & 67 & $93_{-76}^{+283}$ & / & \citetalias{David2015} \\
\noalign{\vskip 1mm}
HIP7345 & 01 34 37.7 & -15 40 34.89  &94.84 & -3.14 & 5.5 & A1V & 61 & $35_{-5}^{+5}$ & Y & \citetalias{Zuckerman1995,Zuckerman2012,Galicher2016} \\
\noalign{\vskip 1mm}
HIP7805 & 01 40 24.0 & -60 59 53.62  & 61.94 & -10.50 & 6.7 & F2V & 66 & $30_{-15}^{+15}$ & Y & \citetalias{Zuckerman2001,Zuckerman2004,Meshkat2017} \\
\noalign{\vskip 1mm}
HIP8832 & 01 53 31.8 & +19 17 37.87  & 79.20 & -97.63 & 2.8 & A0 & 50 & $87_{-71}^{+195}$ & / & \citetalias{David2015} \\
\noalign{\vskip 1mm}

HIP9902 & 02 07 26.1 & -59 40 45.942  & 91.11 & -18.29 & 6.2 & F7V & 44 & $45_{-4}^{+4}$ & Y & \citetalias{Kennedy2014,Bell2015} \\
\noalign{\vskip 1mm}
HIP13141 & 02 49 01.4 & -62 48 23.47  & 94.02 & 29.10 & 5.2 & A2V & 50 & $100_{-70}^{+200}$ & Y & \citetalias{Rhee2007,Galicher2016} \\
\noalign{\vskip 1mm}

HIP16095 & 03 27 18.6 & +12 44 07.03  & 10.36 & -7.56 & 6.3 & A0V & 88 & $194_{-138}^{+171}$ & / & \citetalias{Zuckerman2013,David2015} \\
\noalign{\vskip 1mm}

HIP18437 & 03 56 29.3 & -38 57 43.80  & 29.46 & 0.10 & 6.8 & A0V & 100 & $187_{-177}^{+150}$ & Y & \citetalias{Rhee2007,Meshkat2017} \\
\noalign{\vskip 1mm}

HIP19990 & 04 17 15.6 & +20 34 42.93  & -39.41 & -60.79 & 4.6 & A3 & 29 & $70_{-40}^{+30}$ & / & \citetalias{Zuckerman2013,Galicher2016} \\
\noalign{\vskip 1mm}

HIP22192 & 04 46 25.7 & -28 05 14.8  & -3.82 & 17.58 & 5.7 & A3V & 56 & $12_{-5}^{+5}$ & / & \citetalias{Zuckerman2013,Galicher2016} \\
\noalign{\vskip 1mm}

HIP22226 & 04 46 49.5 & -26 18 08.84  & 34.52 & -4.13 & 6.9 & F3V  & 78 & $30_{-20}^{+20}$ & Y & \citetalias{Rhee2007,Galicher2016} \\
\noalign{\vskip 1mm}

HIP22845 & 04 54 53.7 & +10 09 02.99  & 41.49 & -128.73 & 4.5 & A3V & 34 & $100_{-70}^{+200}$ & Y & \citetalias{Zuckerman2004b,Galicher2016} \\
\noalign{\vskip 1mm}

HIP26309 & 05 36 10.2 & -28 42 28.847  & 25.80 & -3.04 & 5.9 & A2V & 56 & $30_{-10}^{+20}$ & / & \citetalias{Zuckerman2011,Galicher2016} \\
\noalign{\vskip 1mm}

HIP26990 & 05 43 35.8 & -39 55 24.7145  & 25.82 & 15.08 & 6.8 & G0V & 55 & $42_{-7}^{+8}$ & Y & \citetalias{Moor2016,Vigan2017} \\
\noalign{\vskip 1mm}

HIP34276 & 07 06 20.9 & -43 36 38.69  & 5.80 & 13.20 & 6.5 & A0V & 102 & $185_{-170}^{+120}$ & Y & \citetalias{Rhee2007,Meshkat2017} \\
\noalign{\vskip 1mm}

HIP41307 & 08 25 39.6 & -03 54 23.11  & -66.43 & -23.41 & 3.9 & A0V & 37 & $203_{-100}^{+100}$ & Y & \citetalias{Rhee2007,Meshkat2017} \\
\noalign{\vskip 1mm}

HIP93542 & 19 03 06.8 & -42 05 42.38  & 56.41 & -46.43 & 5.0 & B9V & 59  & $76_{-62}^{+148}$ & Y & \citetalias{Rhee2007,David2015} \\
\noalign{\vskip 1mm}
HIP95619 & 19 26 56.4 & -29 44 35.617  & 18.63 & -50.13 & 5.7 & B8.5 & 70 & $86_{-69}^{+138}$ & Y & \citetalias{David2015} \\
\noalign{\vskip 1mm}
HIP97749 & 19 51 50.6 & -39 52 27.7  & 18.42 & -11.27 & 5.4 & A & 100 & $82_{-67}^{+177}$ & / & \citetalias{David2015} \\
\noalign{\vskip 1mm}

HIP101800 & 20 37 49.1 & +11 22 39.63  & 39.15 & -8.26 & 5.4 & A1V & 57 & $225_{-43}^{+311}$ & Y & \citetalias{Rhee2007,David2015} \\
\noalign{\vskip 1mm}
HIP101958 & 20 39 38.2 & +15 54 43.46  & 53.82 & 8.47 & 3.9 & B9V & 77 & $60_{-49}^{+164}$ & / & \citetalias{David2015} \\
\noalign{\vskip 1mm}

HIP117452 & 23 48 55.5 & -28 07 48.97  & 100.80 & -105.34 & 4.6 & A0V & 42 & $70_{-40}^{+30}$ & Y & \citetalias{Zuckerman2011,David2015} \\
\noalign{\vskip 1mm}
\hline
\end{tabular}

\label{table_p99}
\end{table*}

\section{Observations}
\label{sec:observations}

The SPHERE planet-finder instrument installed at the VLT \citep{Beuzit2019} is a highly specialized instrument, dedicated to high-contrast imaging and spectroscopy of young giant exoplanets. It is based on the SAXO extreme adaptive optics (XAO) system \citep{Fusco2006,Sauvage2010,Petit2014}, which controls a deformable mirror with $41\times41$ actuators, and four control loops (fast visible tip-tilt, high-orders, near-infrared differential tip-tilt, and pupil stabilization).  The common path optics employ several stress-polished toric mirrors \citep{Hugot2012} to transport the beam to the coronagraphs and scientific instruments. Several types of coronagraphic devices for stellar diffraction suppression are provided, including apodized pupil Lyot coronagraphs \citep{Soummer2005} and achromatic four-quadrant phase masks \citep{Boccaletti2008}. The instrument has three science subsystems: the infrared dual-band imager and spectrograph (IRDIS, \citealt{Dohlen2008}), an integral field spectrograph (IFS; \citealt{Claudi2008}), and the Zimpol rapid-switching imaging polarimeter (ZIMPOL; \citealt{Thalmann2008}).\\

The sample of young early-type stars was observed using the IRDIFS-EXT mode, with IRDIS in the dual-band imaging (DBI, \citealt{Vigan2010}) mode with $K_1K_2$ filters ($\lambda_{K_1} = 2.1025 \pm 0.1020,\mu$m - $\lambda_{K_2} = 2.2550 \pm 0.1090,\mu$m), and IFS in the $Y-H$ ($0.97-1.66\,\mu$m) mode in pupil-tracking. This combination enables the use of angular and/or spectral differential imaging techniques to improve the contrast performances at the subarcsecond level \citep{Racine1999,Marois2006}. The choice between IRDIFS mode and IRDIFS-EXT mode is critical to optimizing the detection of young, early-T, or warm, mid-L dwarfs planets, considering the primary age and distance. Indeed, it was crucial in the cases of the $\beta$ Pic\,b \citep{Lagrange2009} and HD\,95086\,b \citep{Rameau2013} discoveries to properly remove quasi-static speckles that dominate performance detection at close inner angles ($0.1-2.0\,\!''$, i.e., $3-60$\,au at 30 pc), but to also maximize the emitted flux by the giant planets. For young ages ($10-50$\,Myr), as the potential planets to which we are mostly sensitive are warm and dusty L-type planets with no methane absorption, the choice of the IRDIFS-EXT mode is more appropriate and was chosen for this observing campaign. For the follow-up, as candidates were only detected in the IRDIS field of view, we opted for the use of the IRDIS the DBI mode with $J_2J_3$ filters ($\lambda_{J_2} =2.1025 \pm 0.1020\,\mu$m - $\lambda_{J_3} = 2.2550 \pm 0.1090\,\mu$m) in pupil-tracking. Thus, this second epoch provides, in addition to the possibility of checking for common proper motion of the candidates relative to the primary star, the possibility to better discriminate background stars from physically young, early-T, or warm mid-L dwarfs planets in the color-magnitude diagram \citep{Bonnefoy2018}.
\begin{table*}
\caption{Observing Log}
\begin{center}
\begin{tabular}{lllllllllll}
\hline 
\hline
UT Date & Target & Instrument  & Mode  & Filter  & NDIT $\times$ DIT  & $N_{\rm{exp}}$  & $\Delta \theta$ & $\omega$ & Strehl & Airmass  \\
 &&&&& (s) && ($^o$) & (") & $@1.6 \upmu \mathrm{m}$ \\
\hline
\multicolumn{11}{c}{Survey}\\
\hline 
\multirow{4}{*}{05-10-2016} & \multirow{2}{*}{HIP9902}  & IRDIS & DBI & $K_1$$K_2$ & $3 \times 64$ & \multirow{2}{*}{46} & \multirow{2}{*}{20.7} & \multirow{2}{*}{0.62}  & \multirow{2}{*}{0.75} & \multirow{2}{*}{1.22} \\
&& IFS & $R_{\lambda} = 30$ & $$YJH$$ & $1 \times 64$ &  \\

& \multirow{2}{*}{HIP18437} & IRDIS & DBI & $K_1$$K_2$ & $3 \times 64$ & \multirow{2}{*}{46} & \multirow{2}{*}{44.2} & \multirow{2}{*}{0.47}  & \multirow{2}{*}{0.77} & \multirow{2}{*}{1.03} \\
&& IFS & $R_{\lambda} = 30$ & $$YJH$$ & $1 \times 64$ &  \\

\multirow{4}{*}{07-10-2016} & \multirow{2}{*}{HIP7805}  & IRDIS & DBI & $K_1$$K_2$ & $3 \times 64$ & \multirow{2}{*}{46} & \multirow{2}{*}{20.0} & \multirow{2}{*}{0.53}  & \multirow{2}{*}{0.83} & \multirow{2}{*}{1.24} \\
&& IFS & $R_{\lambda} = 30$ & $$YJH$$ & $1 \times 64$ &  \\

& \multirow{2}{*}{HIP16095} & IRDIS & DBI & $K_1$$K_2$ & $3 \times 64$ & \multirow{2}{*}{46} & \multirow{2}{*}{19.0} & \multirow{2}{*}{0.46}  & \multirow{2}{*}{0.87} & \multirow{2}{*}{1.26} \\
&& IFS & $R_{\lambda} = 30$ & $$YJH$$ & $1 \times 64$ &  \\

\multirow{2}{*}{08-10-2016} & \multirow{2}{*}{HIP13141}  & IRDIS & DBI & $K_1$$K_2$ & $3 \times 64$ & \multirow{2}{*}{46} & \multirow{2}{*}{20.8} & \multirow{2}{*}{0.41}  & \multirow{2}{*}{0.83} & \multirow{2}{*}{1.30} \\
&& IFS & $R_{\lambda} = 30$ & $$YJH$$ & $1 \times 64$ &  \\

\multirow{2}{*}{10-11-2016} & \multirow{2}{*}{HIP19990}  & IRDIS & DBI & $K_1$$K_2$ & $3 \times 64$ & \multirow{2}{*}{46} & \multirow{2}{*}{22.6} & \multirow{2}{*}{0.27}  & \multirow{2}{*}{0.94} & \multirow{2}{*}{1.30} \\
&& IFS & $R_{\lambda} = 30$ & $$YJH$$ & $1 \times 32$ &  \\

\multirow{2}{*}{12-11-2016} & \multirow{2}{*}{HIP26309}  & IRDIS & DBI & $K_1$$K_2$ & $3 \times 64$ & \multirow{2}{*}{46} & \multirow{2}{*}{107.4} & \multirow{2}{*}{0.41}  & \multirow{2}{*}{0.87} & \multirow{2}{*}{1.01} \\
&& IFS & $R_{\lambda} = 30$ & $$YJH$$ & $1 \times 64$ &  \\

\multirow{2}{*}{13-11-2016} & \multirow{2}{*}{HIP22192}  & IRDIS & DBI & $K_1$$K_2$ & $7 \times 32$ & \multirow{2}{*}{46} & \multirow{2}{*}{130.9} & \multirow{2}{*}{0.33}  & \multirow{2}{*}{0.86} & \multirow{2}{*}{1.01} \\
&& IFS & $R_{\lambda} = 30$ & $$YJH$$ & $1 \times 32$ &  \\

\multirow{2}{*}{04-12-2016} & \multirow{2}{*}{HIP7345}  & IRDIS & DBI & $K_1$$K_2$ & $3 \times 64$ & \multirow{2}{*}{17} & \multirow{2}{*}{81.4} & \multirow{2}{*}{0.44}  & \multirow{2}{*}{0.90} & \multirow{2}{*}{1.02} \\
&& IFS & $R_{\lambda} = 30$ & $$YJH$$ & $1 \times 64$ &  \\

\multirow{2}{*}{05-12-2016} & \multirow{2}{*}{HIP22226}  & IRDIS & DBI & $K_1$$K_2$ & $3 \times 64$ & \multirow{2}{*}{46} & \multirow{2}{*}{15.2} & \multirow{2}{*}{0.42}  & \multirow{2}{*}{0.82} & \multirow{2}{*}{1.00} \\
&& IFS & $R_{\lambda} = 30$ & $$YJH$$ & $1 \times 64$ &  \\

\multirow{2}{*}{07-12-2016} & \multirow{2}{*}{HIP22845}  & IRDIS & DBI & $K_1$$K_2$ & $3 \times 64$ & \multirow{2}{*}{46} & \multirow{2}{*}{19.3} & \multirow{2}{*}{0.44}  & \multirow{2}{*}{0.82} & \multirow{2}{*}{1.27} \\
&& IFS & $R_{\lambda} = 30$ & $$YJH$$ & $1 \times 32$ &  \\

\multirow{2}{*}{13-12-2016} & \multirow{2}{*}{HIP34276}  & IRDIS & DBI & $K_1$$K_2$ & $8 \times 32$ & \multirow{2}{*}{46} & \multirow{2}{*}{39.5} & \multirow{2}{*}{0.55}  & \multirow{2}{*}{0.84} & \multirow{2}{*}{1.06} \\
&& IFS & $R_{\lambda} = 30$ & $$YJH$$ & $1 \times 64$ &  \\

\multirow{4}{*}{15-12-2016} & \multirow{2}{*}{HIP26990}  & IRDIS & DBI & $K_1$$K_2$ & $3 \times 64$ & \multirow{2}{*}{46} & \multirow{2}{*}{42.6} & \multirow{2}{*}{0.55}  & \multirow{2}{*}{0.76} & \multirow{2}{*}{1.04} \\
&& IFS & $R_{\lambda} = 30$ & $$YJH$$ & $1 \times 64$ &  \\

& \multirow{2}{*}{HIP41307} & IRDIS & DBI & $K_1$$K_2$ & $17 \times 16$ & \multirow{2}{*}{46} & \multirow{2}{*}{43.0} & \multirow{2}{*}{0.35}  & \multirow{2}{*}{0.92} & \multirow{2}{*}{1.03} \\
&& IFS & $R_{\lambda} = 30$ & $$YJH$$ & $1 \times 16$ &  \\ 

\multirow{4}{*}{17-06-2017} & \multirow{2}{*}{HIP93542}  & IRDIS & DBI & $K_1$$K_2$ & $7 \times 32$ & \multirow{2}{*}{46} & \multirow{2}{*}{59.5} & \multirow{2}{*}{0.83}  & \multirow{2}{*}{0.69} & \multirow{2}{*}{1.05} \\
&& IFS & $R_{\lambda} = 30$ & $$YJH$$ & $1 \times 32$ &  \\

& \multirow{2}{*}{HIP97749} & IRDIS & DBI & $K_1$$K_2$ & $7 \times 32$ & \multirow{2}{*}{46} & \multirow{2}{*}{43.3} & \multirow{2}{*}{0.81}  & \multirow{2}{*}{0.52} & \multirow{2}{*}{1.06} \\
&& IFS & $R_{\lambda} = 30$ & $$YJH$$ & $1 \times 32$ &  \\

\multirow{2}{*}{06-07-2017} & \multirow{2}{*}{HIP101800}  & IRDIS & DBI & $K_1$$K_2$ & $7 \times 32$ & \multirow{2}{*}{42} & \multirow{2}{*}{22.1} & \multirow{2}{*}{0.58}  & \multirow{2}{*}{0.86} & \multirow{2}{*}{1.24} \\
&& IFS & $R_{\lambda} = 30$ & $$YJH$$ & $1 \times 32$ &  \\

\multirow{2}{*}{15-07-2017} & \multirow{2}{*}{HIP117452}  & IRDIS & DBI & $K_1$$K_2$ & $6 \times 32$ & \multirow{2}{*}{46} & \multirow{2}{*}{117.1} & \multirow{2}{*}{0.45}  & \multirow{2}{*}{0.87} & \multirow{2}{*}{1.01} \\
&& IFS & $R_{\lambda} = 30$ & $$YJH$$ & $1 \times 32$ &  \\

\multirow{2}{*}{20-07-2017} & \multirow{2}{*}{HIP101958}  & IRDIS & DBI & $K_1$$K_2$ & $15 \times 16$ & \multirow{2}{*}{46} & \multirow{2}{*}{23.4} & \multirow{2}{*}{0.45}  & \multirow{2}{*}{0.90} & \multirow{2}{*}{1.36} \\
&& IFS & $R_{\lambda} = 30$ & $$YJH$$ & $1 \times 16$ &  \\

\multirow{2}{*}{31-07-2017} & \multirow{2}{*}{HIP95619}  & IRDIS & DBI & $K_1$$K_2$ & $7 \times 32$ & \multirow{2}{*}{46} & \multirow{2}{*}{110.0} & \multirow{2}{*}{0.77}  & \multirow{2}{*}{0.62} & \multirow{2}{*}{1.01} \\
&& IFS & $R_{\lambda} = 30$ & $$YJH$$ & $1 \times 32$ &  \\

\multirow{2}{*}{09-08-2017} & \multirow{2}{*}{HIP8832}  & IRDIS & DBI & $K_1$$K_2$ & $15 \times 16$ & \multirow{2}{*}{46} & \multirow{2}{*}{22.5} & \multirow{2}{*}{0.35}  & \multirow{2}{*}{0.89} & \multirow{2}{*}{1.40} \\
&& IFS & $R_{\lambda} = 30$ & $$YJH$$ & $1 \times 16$ &  \\

\multirow{2}{*}{10-09-2017} & \multirow{2}{*}{HIP3277}  & IRDIS & DBI & $K_1$$K_2$ & $7 \times 32$ & \multirow{2}{*}{46} & \multirow{2}{*}{26.5} & \multirow{2}{*}{0.54}  & \multirow{2}{*}{0.83} & \multirow{2}{*}{1.20} \\
&& IFS & $R_{\lambda} = 30$ & $$YJH$$ & $1 \times 32$ &  \\

\hline   
\multicolumn{11}{c}{Follow-up}\\
\hline   
\multirow{1}{*}{27-09-2018} & \multirow{1}{*}{HIP117452}  & IRDIS & DBI & $J_2$$J_3$ & $ 6 \times 32 $ & \multirow{1}{*}{22} & \multirow{1}{*}{112.7} & \multirow{1}{*}{0.41}  & \multirow{1}{*}{0.88} & \multirow{1}{*}{1.00}  \\

\multirow{1}{*}{10-10-2018} & \multirow{1}{*}{HIP8832}  & IRDIS & DBI & $J_2$$J_3$ & $ 4 \times 48 $ & \multirow{1}{*}{23} & \multirow{1}{*}{20.4} & \multirow{1}{*}{0.61}  & \multirow{1}{*}{0.78} & \multirow{1}{*}{1.00}  \\

\multirow{1}{*}{22-11-2018} & \multirow{1}{*}{HIP34276}  & IRDIS & DBI & $J_2$$J_3$ & $ 4 \times 64 $ & \multirow{1}{*}{23} & \multirow{1}{*}{46.3} & \multirow{1}{*}{0.39}  & \multirow{1}{*}{0.82} & \multirow{1}{*}{1.44}  \\

\multirow{1}{*}{09-05-2019} & \multirow{1}{*}{HIP95619}  & IRDIS & DBI & $J_2$$J_3$ & $ 7 \times 32 $ & \multirow{1}{*}{23} & \multirow{1}{*}{22.3} & \multirow{1}{*}{0.51}  & \multirow{1}{*}{0.75} & \multirow{1}{*}{1.02}  \\

\multirow{1}{*}{18-06-2019} & \multirow{1}{*}{HIP101800}  & IRDIS & DBI & $J_2$$J_3$ & $ 7 \times 32 $ & \multirow{1}{*}{23} & \multirow{1}{*}{20.2} & \multirow{1}{*}{0.68}  & \multirow{1}{*}{0.83} & \multirow{1}{*}{1.36}  \\

\hline   
\end{tabular}
\end{center}
\label{table_obs_log}
\end{table*}

The observing sequence used for the survey is as follows; PSF flux reference, coronographic centering using the waffle spots, deep coronographic observation of about 70\,min in total on target, new coronographic centering using the waffle spots, PSF flux reference, and sky. The PSF flux references were used to estimate the relative photometry of the companion candidates detected in the IRDIS and IFS field of view, as well as the detection limits. The coronographic centering sequence using the waffle spots sequence is critical to obtaining the position of the star behind the coronograph and the relative astrometry of the companion candidates. The deep coronographic observation was obtained close to meridian to maximize the field rotation. Finally, the sky background was used to optimize the background subtraction and the flat field correction. The typical observing sequence lasts approximately 90\,min, including pointing and overheads. The detail of the observations per target is reported in Table\,\ref{table_obs_log}. As a by-product of the SPHERE observation, one can access the evolution of the different atmospheric parameters seen and registered by the SPHERE XAO system (SAXO). These real-time parameters are good diagnostics of the turbulence conditions ($\tau_0$, $r_0$, integrated wind over the line of sight) and of the XAO correction (Strehl at 1.6\,$\mu$m) during the observing sequence. The summary of these SAXO parameters over the full survey is reported in Table\,\ref{table_obs_log} and shown in Figure\,\ref{sparta}. Given the brightness of our targets, about $70\,\%$ of the survey was obtained under median or good conditions for Paranal, with a typical Strehl ratio larger than $80\,\%$. Prior to the UT3 intervention at VLT in 2017, a few cases were affected by the low-wind effect, despite good atmospheric conditions.

\begin{figure*}[t]
\centering
\includegraphics[width=\columnwidth]{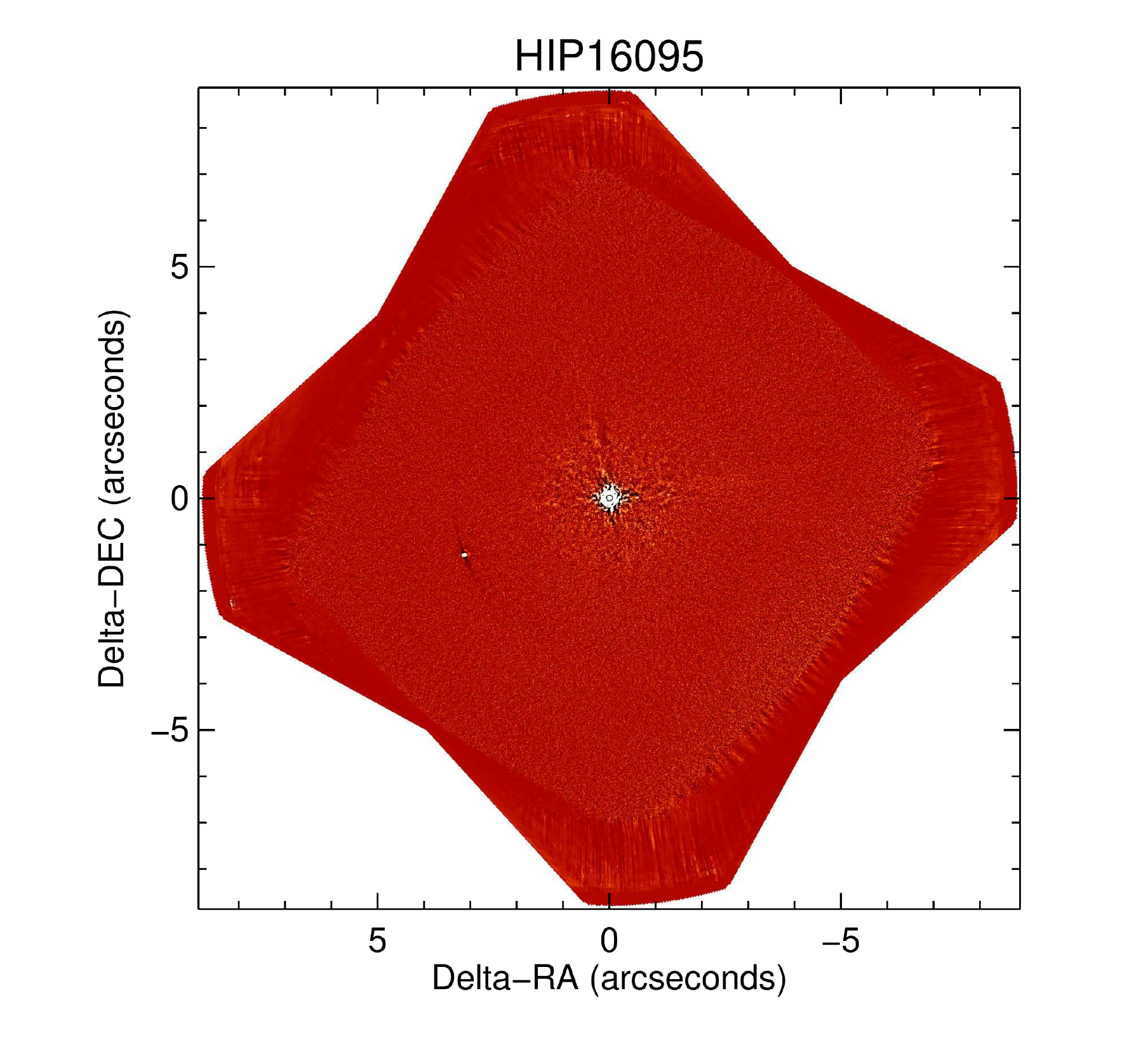}
\includegraphics[width=\columnwidth]{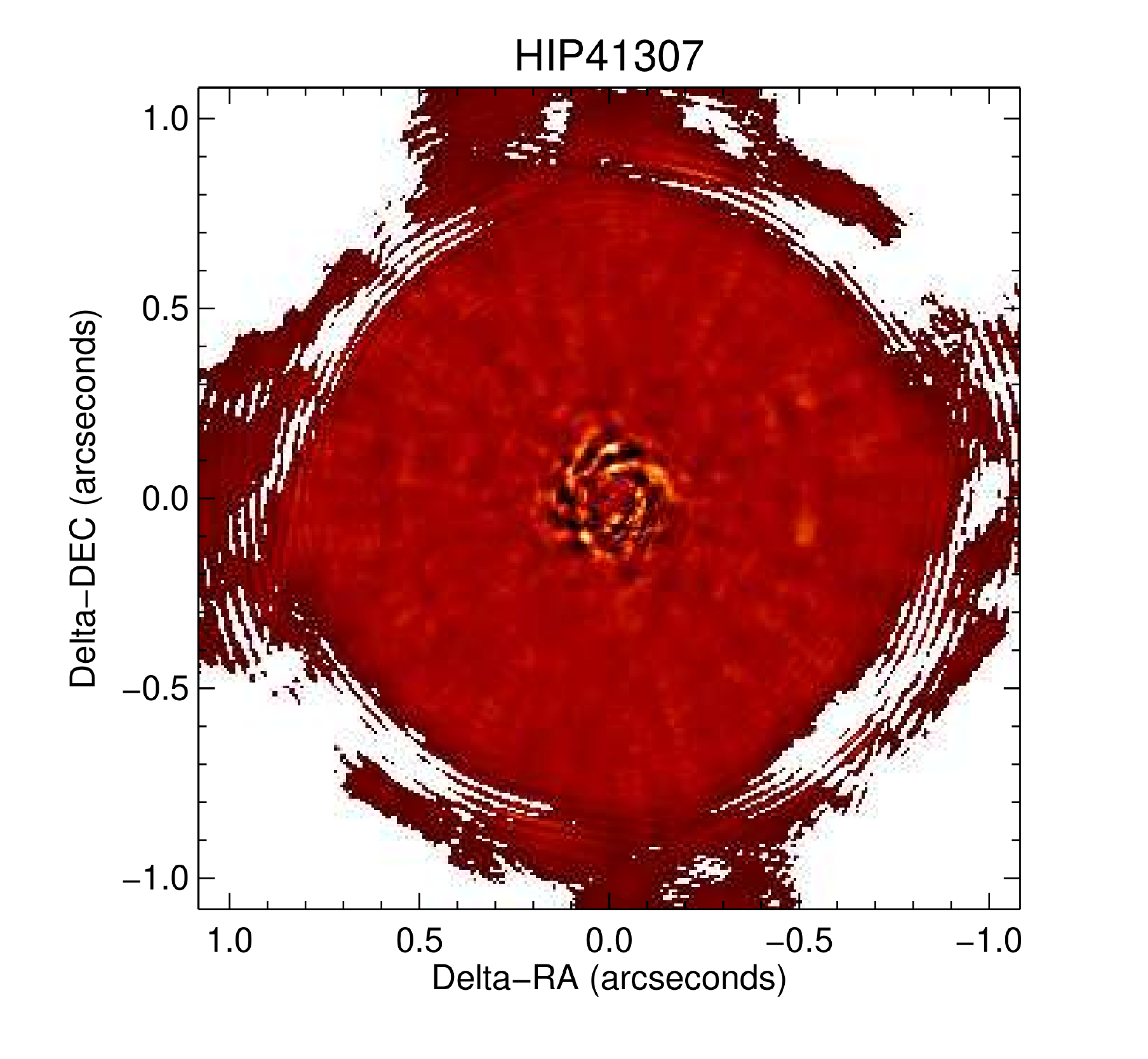}
\caption{Left: IRDIS reduced full-frame image of HIP\,16095 in the $K_1$ and $K_2$  combined filters using SpeCal with the TLOCI algorithm \citep{Galicher2011}. A bright companion candidate is well identified at a few arcseconds to the east of the star. North is up, and east is left. Right: IFS image reduced in PCA ASDI of HIP\,41307.
}
\label{obj_hip16095}
\end{figure*}

\section{Data reduction and analysis}
\label{sec:data_reduc_analysis}

In order to calibrate the IRDIS and IFS dataset on sky, the platescale and true north solution at each epoch were corrected based on the long-term analysis of the SPHERE Guaranteed Time Observation astrometric calibration described by \cite{Maire2016}. The rotation correction considered to align images to the detector vertical in pupil-tracking observations is $-135.99\pm0.11\degr$. Anamorphism correction was obtained by stretching the image $Y$-direction with a factor of $1.0060\pm0.0002$. All IRDIS and IFS datasets were reduced using the SPHERE Data Reduction and Handling (DRH) automated pipeline \citep{Pavlov2008}  and additional IDL routines for the IFS data reduction \citep{Mesa2015} at the SPHERE Data Center \citep{Delorme2017} to correct each data cube for bad pixels, dark current, flat field, and sky background. After combining all data cubes with an adequate calculation of the parallactic angle for each individual frame of the deep coronagraphic sequence, all frames were shifted at the position of the stellar centroid calculated from the initial star center position.\\

For an independent check, two pipelines were then used to process the data in angular differential imaging (ADI), and in combined spectral and angular differential imaging (ASDI): the IPAG-ADI pipeline \citep{Chauvin2012}, and the SpeCal \citep{Galicher2018}. These routines allowed us to reduce the data cubes with almost the same set of algorithms (classical ADI, \citealt{Marois2006}; LOCI, \citealt{Lafreniere2007}; PCA,\citealt{Soummer2012}; Andromeda, \citealt{Cantalloube2015}), and to exploit the spectral diversity of the IRDIS and IFS observations using ASDI techniques in addition to ADI only. Following the principles described in \cite{Galicher2018}, SpeCal (and IPAG-ADI) delivers, for various algorithms and observing techniques (ADI, ASDI), contrast curves, signal-to-noise ratio (S/N)  maps, and the possibility to locally characterize the astrometric, photometric, and spectroscopic signal of any companion candidate using either a template or negative fake planet injection approach. As consistent results were found with both pipelines, the full set of observations was reduced with SpeCal (routinely used with the SPHERE GTO) using the TLOCI algorithm (in ADI and ASDI) for IRDIS, and the PCA algorithm (in ASDI) for IFS. A spatial filtering for each data cube was automatically applied to the deep coronographic observations and the reference PSFs before the use of SpeCal. \\ 

The TLOCI algorithm is implemented in SpeCal, to attenuate the background signal. The TLOCI algorithm locally subtracts the stellar speckle pattern for each frame in annuli of $1.5\times\textit{FWHM}$ further divided into sectors. The subtraction is based on a linear combination of the best 20 ($N$ parameter) correlated reference images calculated in the optimization region and selected to minimize the self-subtraction at a maximum of $20\%$ ($\tau$ parameter), see \citealt{Galicher2011} and \citealt{Marois2014} for a further description of the reference frame selection and the subtraction and optimization regions. For IFS, in the PCA version, each frame we used is subtracted from its average over the field of view to estimate the principal components. The spectral diversity is exploited after proper rescaling and renormalization of the IFS data cubes as detailed by \citep{Mesa2015}. Considering the significant field rotation of our observations, the first 100 principal components were subtracted.

\section{Companion candidate detection and characterization}
\label{sec:cc_detection}
 
\begin{figure*}[t]
\centering

\includegraphics[width=0.48\textwidth]{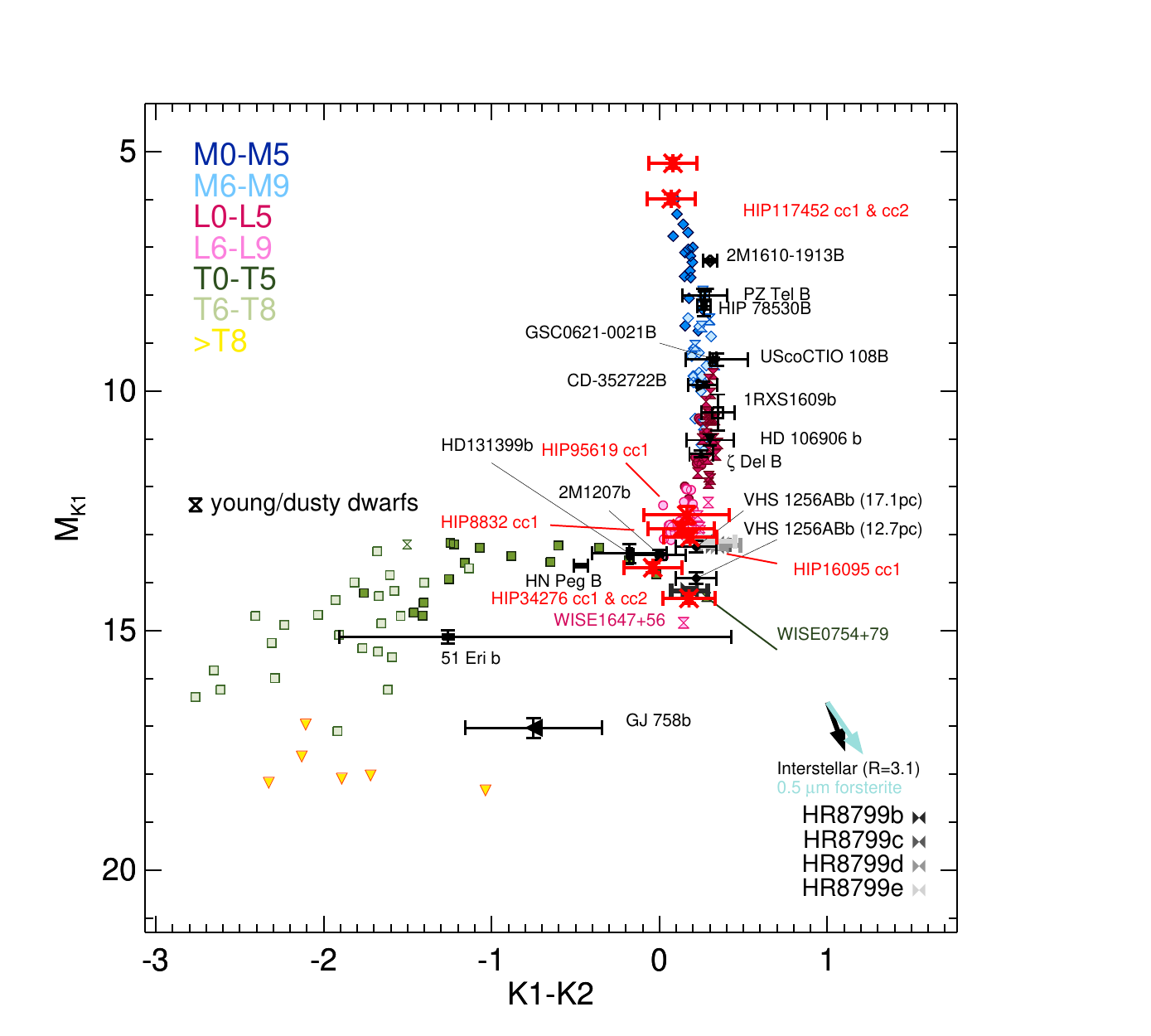}
\includegraphics[width=0.48\textwidth]{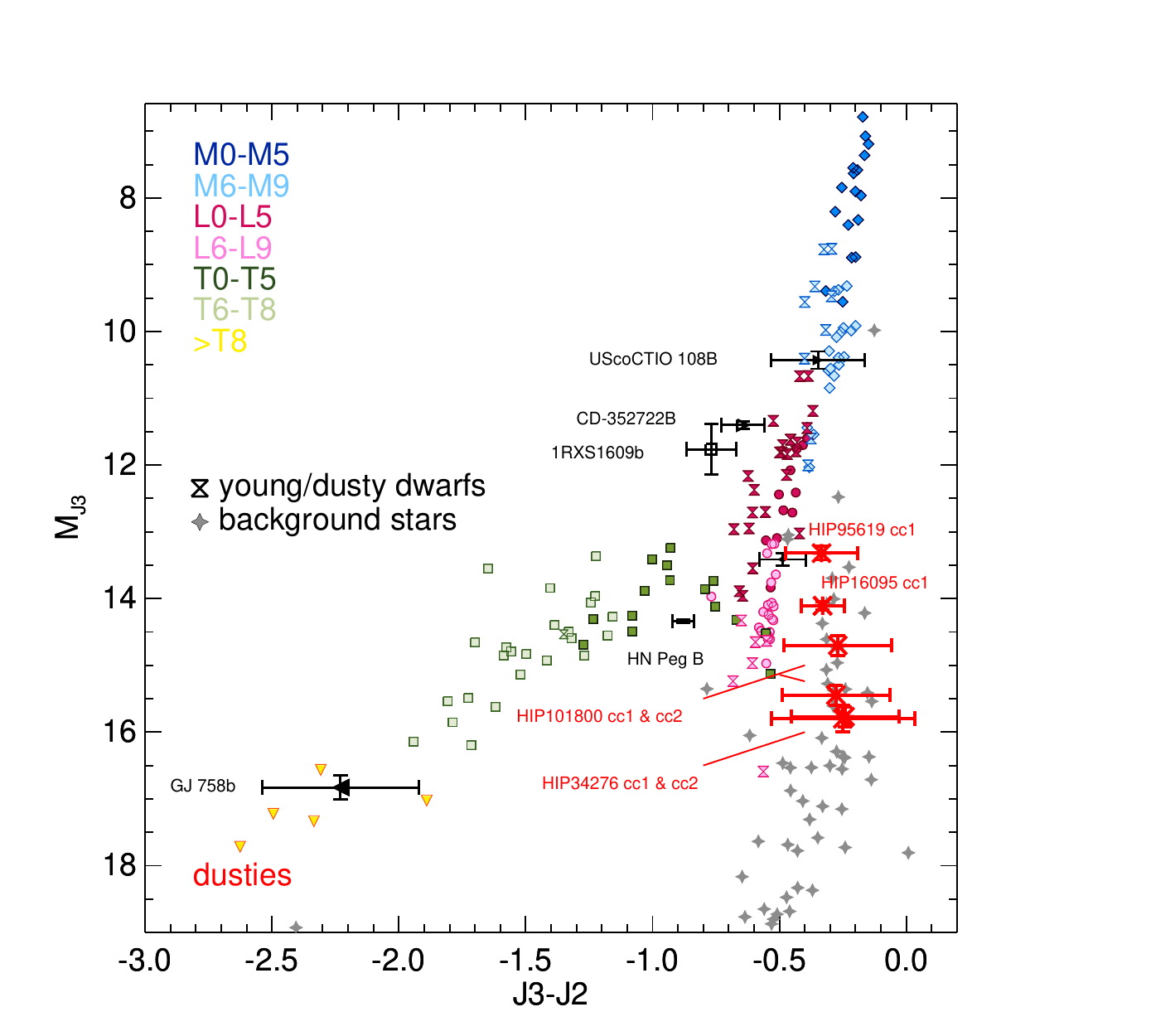}
\caption{Left: absolute magnitude in $K_1$-band versus $K_1$-$K_2$ color for brown dwarfs with discovered companions. Right: same, but for absolute magnitude in $J_3$-band versus $J_3-J_2$ color. The targets from our survey are noted in red.}
\label{diag_mag}
\end{figure*}

\begin{table*}[t]
\caption{Companion candidate characterization and identification. Target name and observing date (modified Julian day) are given, as well as the different sources identified with their relative position and relative flux.}
\begin{center}
\begin{tabular}{llllllll}
\hline 
\hline
Target & UT Date & Candidate  &  Filter &  Separation  & Position angle  & $\Delta_{\rm Filter-1}$ & $\Delta_{\rm Filter-2}$  \\
 & & & &  (mas) & (deg) & (mag) & (mag) \\
\hline
HIP16095 & 57669.2937186 & cc-1 & DK12 & $3368\pm2$ & $111.38\pm0.04$ & $11.46\pm0.12$ & $11.28\pm0.12$ \\
         & 58092.1556576 & cc-1 & DJ23 & $3385\pm2$ & $111.21\pm0.02$ & $12.88\pm0.08$ & $12.55\pm0.09$ \\
HIP95619 & 57965.1627630 & cc-1 & DK12 & $4564\pm3$ & $254.25\pm0.03$ & $11.11\pm0.51$ & $10.94\pm0.54$ \\
         & 58613.3454076 & cc-1 & DJ23 & $4597\pm2$ & $255.23\pm0.01$ & $12.17\pm0.24$ & $11.83\pm0.29$\\
HIP101800&57940.3125070  & cc-1 & DK12 & $4513\pm4$ & $89.84\pm0.037$ & $12.42\pm0.12$ & - \\ 
         &58653.3759935  & cc-1 & DJ23 & $4418\pm2$ & $89.82\pm0.01 $ & $13.34\pm0.17$ & $13.07\pm0.15$ \\ 
         &58653.3759935  & cc-2 & DJ23 & $4021\pm2$ & $89.83\pm0.01 $ & $14.42\pm0.19$ & $14.17\pm0.17$ \\ 
          
HIP34276 & 57736.2557381 & cc-1 & DK12 & $3108\pm7$ & $132.55\pm0.11$ & $12.90\pm0.12$ & $12.72\pm0.13$\\
         & 57736.2557381 & cc-2 & DK12 & $4407\pm4$ & $138.56\pm0.06$ & $12.26\pm0.12$ & $12.30\pm0.12$ \\
         & 58445.3349875 & cc-1 & DJ23 & $3124\pm4$ & $133.01\pm0.06$ & $14.58\pm0.29$ & $14.34\pm0.12$\\ 
         & 58445.3349875 & cc-2 & DJ23 & $4421\pm5$ & $138.95\pm0.06$ & $14.30\pm0.29$ & $14.02\pm0.13$\\ 
HIP117452 & 57949.3975893 & Ba  & DK12 & $3708\pm9$ & $238.09\pm0.15$ & $3.84\pm0.05$  & $3.76\pm0.05$\\
         &                & Bb  & DK12 & $3318\pm10$& $239.13\pm0.17$ & $4.58\pm0.05$  & $4.51\pm0.05$\\
HIP8832  & 57974.3996411 & cc-1 & DK12 & $5674\pm3$ & $213.71\pm0.04$ & $11.47\pm0.50$ & $11.54\pm0.51$ \\
\hline
\end{tabular}
\end{center}
\label{table_candidates}
\end{table*}

Using the IRDIS and IFS S/N maps provided by SpeCal, we identified a total of eight companion candidates by eye at relatively large separation ($\geq\,3.0\,\!''$) in the IRDIS fields of view of six targets (HIP\,16095, HIP\,95619, HIP\,101800, HIP\,34276, HIP\,117452, and HIP\,8832) of the complete survey. One companion candidate was identified at relatively close separation in the IFS field of view of HIP\,41307, but later flagged as a bright quasi-static speckle through various processing tests, and was therefore discarded. \\

Figure \ref{obj_hip16095} shows the IRDIS image reduced in TLOCI ADI of HIP\,16095 (bright companion located at $3.3\,\!''$ in the $K_1$ and $K_2$ combined filters), and the IFS image reduced in PCA ASDI of HIP\,41307 (quasi-static speckle discarded located at $0.5\,\!''$ in the combined \textit{YJH}-bands) as an illustration of the detection process. All companion candidates were then characterized using SpeCal with the TLOCI algorithm in ADI only, and according to a template approach. The relative astrometry and photometry are reported in Table\,\ref{table_candidates}. As first diagnostics, in Figure\,\ref{diag_mag} (\textit{Left}), we reported the location of all our companion candidates in the $K_1$-band- and $K_2$-band-based color-magnitude diagram (CMD).  Details on the diagrams are given in \cite{Mesa2016,Samland2017,Chauvin2018,Bonnefoy2018}. We used the most recent parallaxes of the young objects from \cite{Greco2016}, and added additional companions \citep{Gauza2015,Stone2016,Derosa2014} at the L/T transition. At first glance, we see that all detected companion candidates fall on the expected sequence of possible bound companions from the early-M spectral type for the candidates around HIP\,117452, late-L spectral types for HIP\,95619, HIP\,16095 and HIP\,8832, to early-T for HIP\,34276. The companion around HIP\,101800 was detected only in $K_1$-band during the first epoch. After a verification of the public archive, the companion candidates around HIP\,34276 (cc1 and cc2) and HIP\,101800 (cc1 and cc2) were previously known and characterized as stationary background sources by \cite{Wahhaj2013} as part of the NICI campaign concerning debris disk stars. Both companion candidates around HIP\,117452 were earlier identified by \cite{Derosa2011} in the course of the Volume-limited A-Star (VAST) survey as a candidate binary companion. They were later confirmed by \cite{Matthews2018} as physically bound, confirming that this system was actually a quadruple system with an A0 primary (HIP\,117452\,A), orbited by a close binary pair Ba and Bb also resolved in this survey, and additionally by a K-type star at about $75\,\!"$.

\begin{figure*}[t]
\centering
\includegraphics[width=0.3\textwidth]{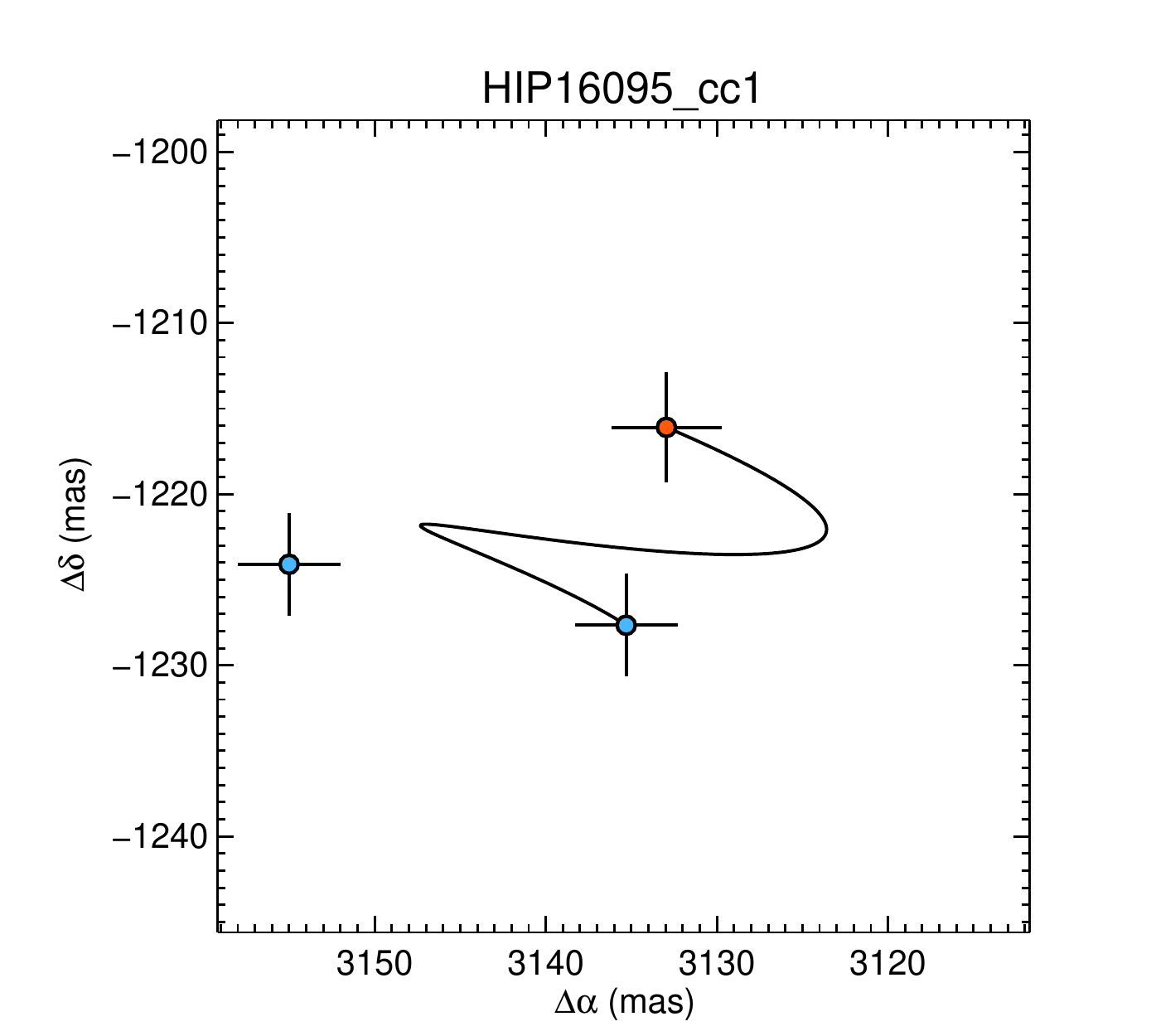}
\includegraphics[width=0.3\textwidth]{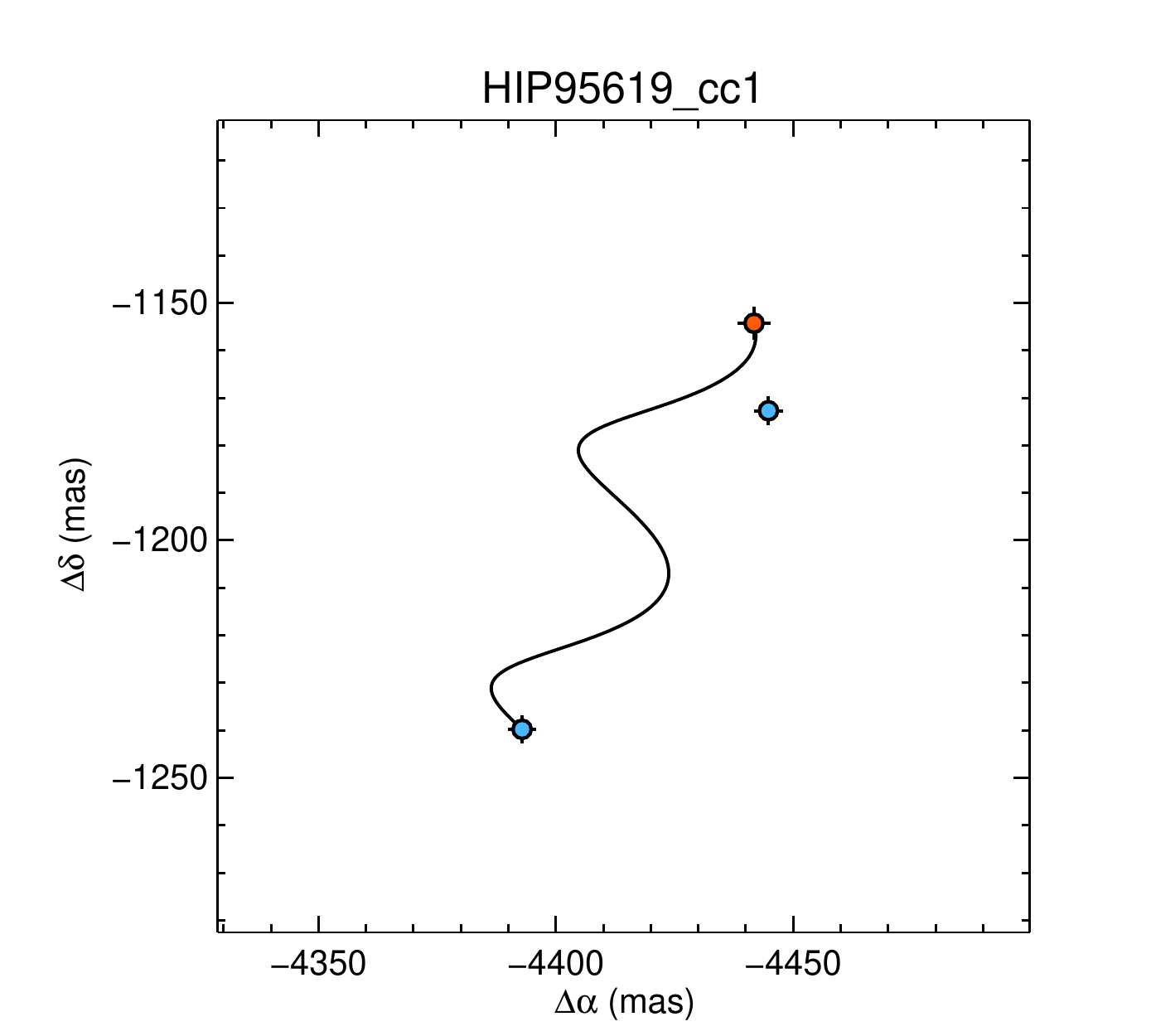}
\includegraphics[width=0.3\textwidth]{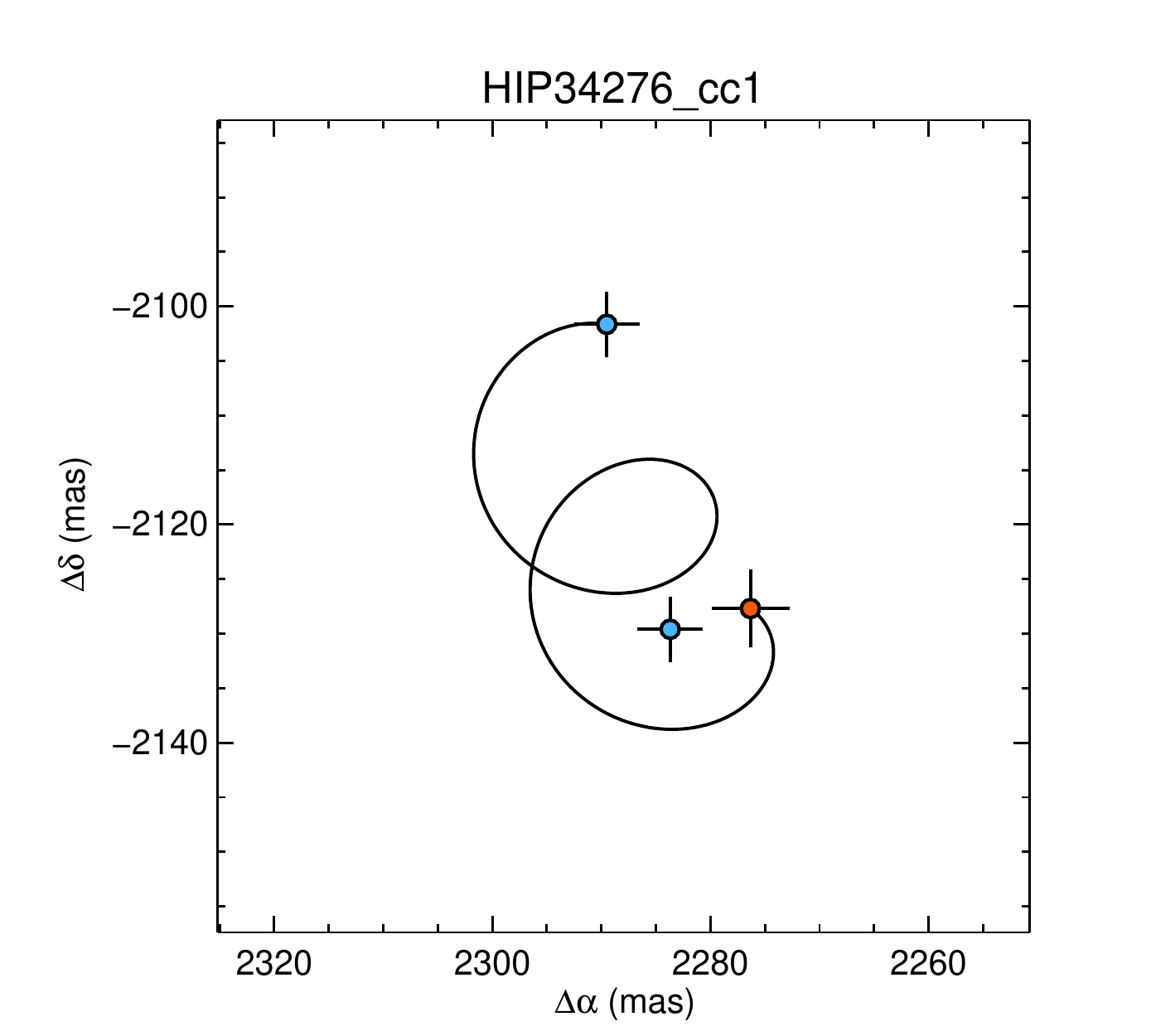}
\includegraphics[width=0.3\textwidth]{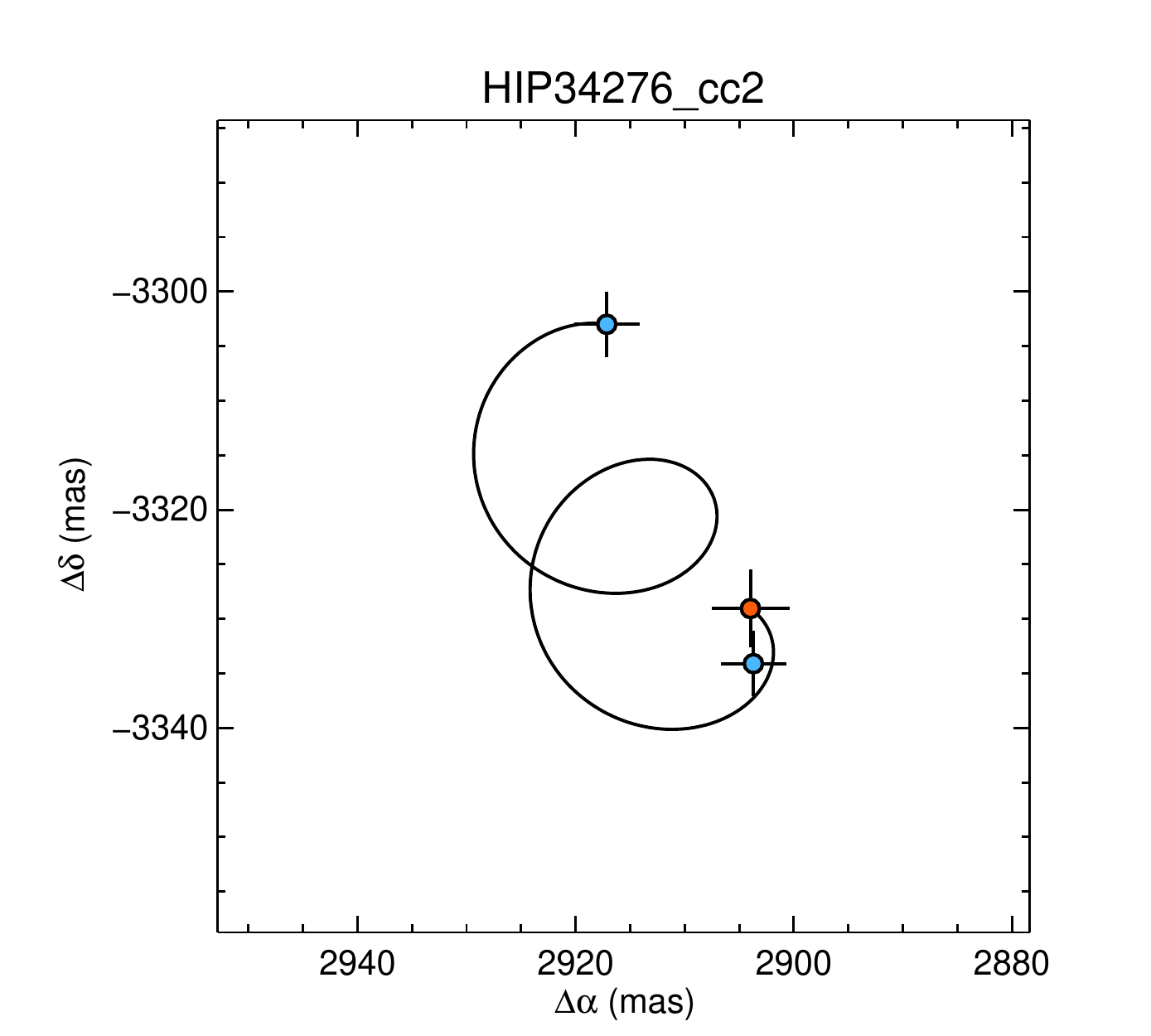}
\includegraphics[width=0.3\textwidth]{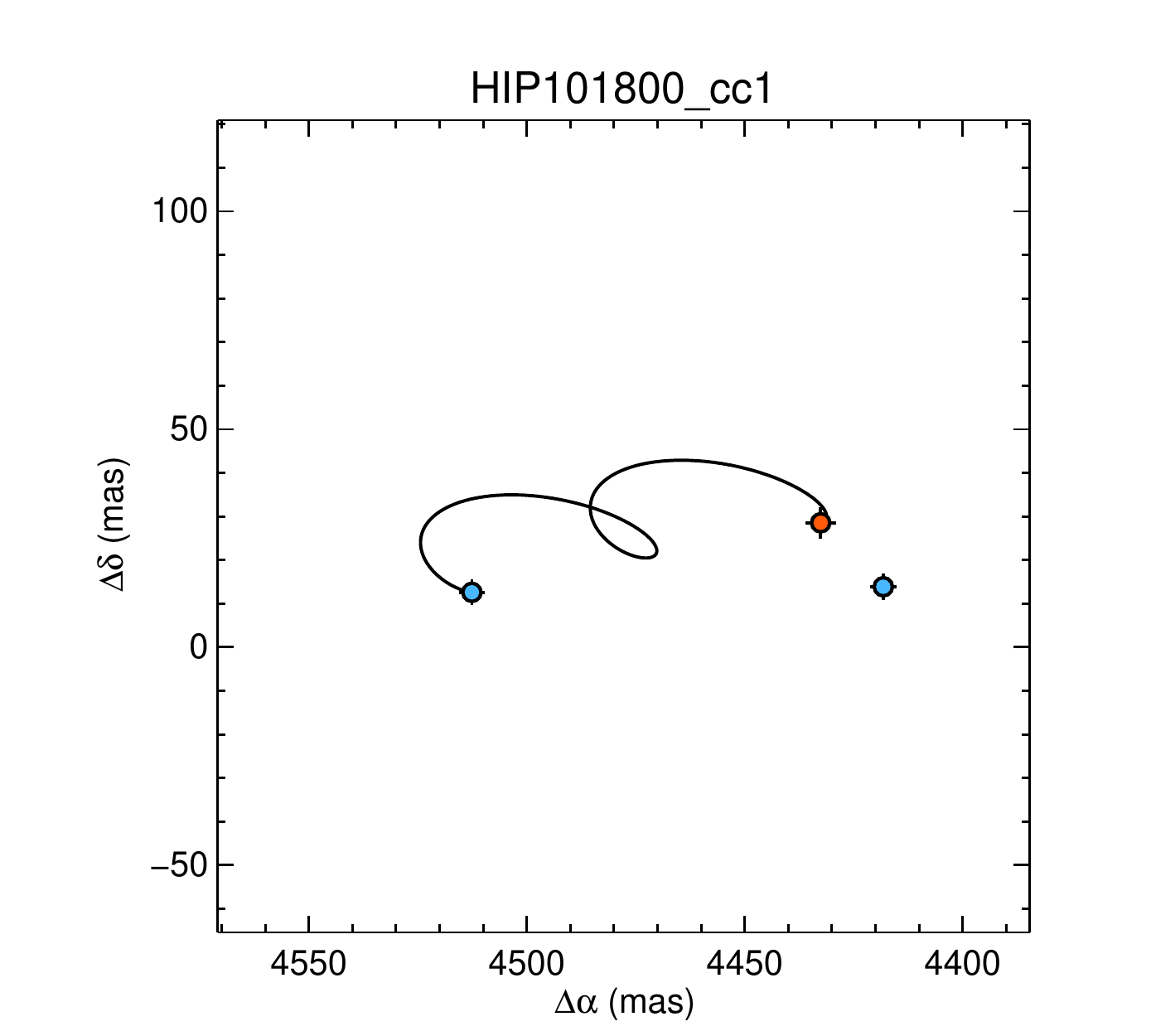}
\includegraphics[width=0.3\textwidth]{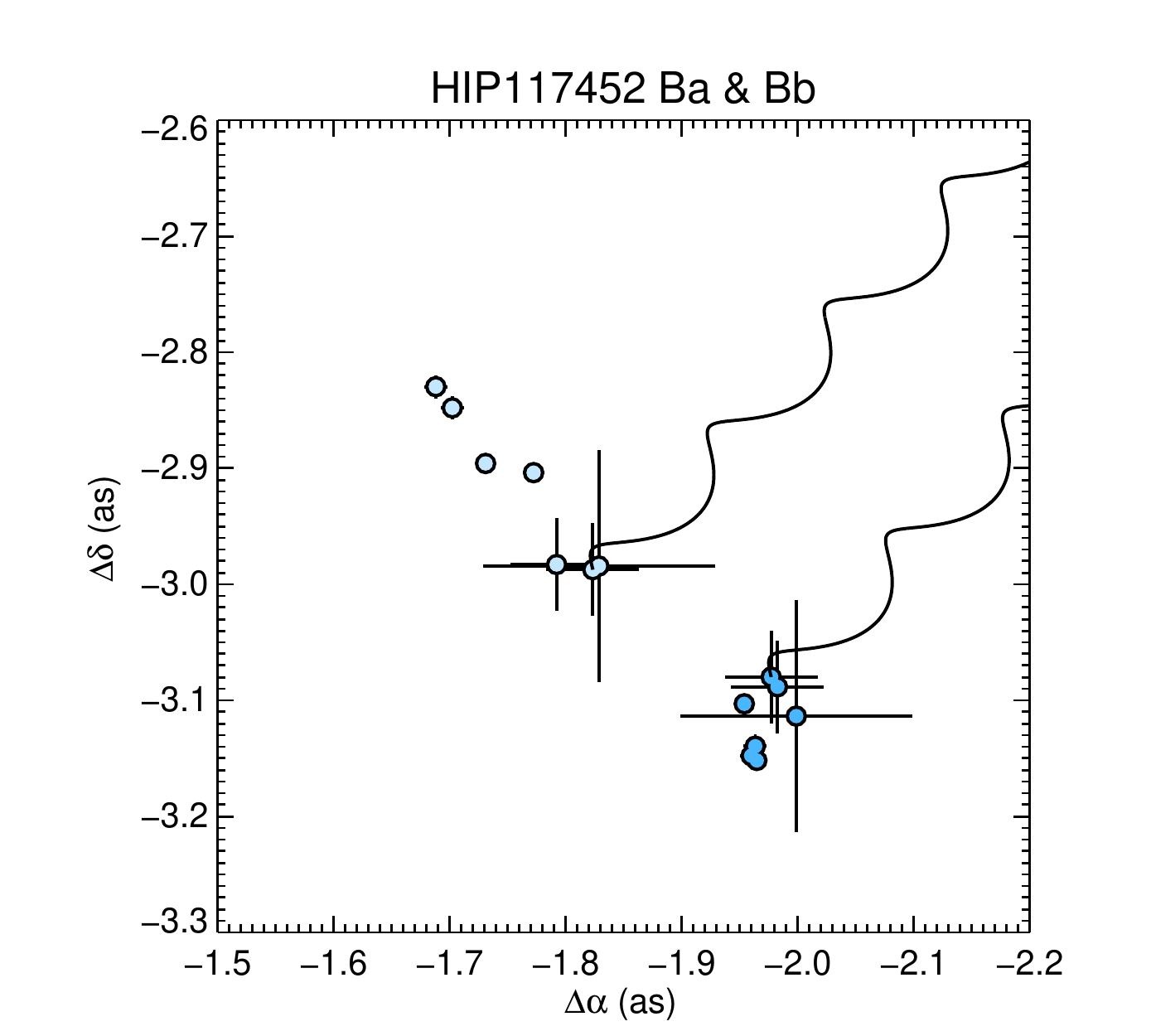}
\caption{SPHERE measurements (in blue) of the offset positions of the companion candidates relative to their primary stars. For each diagram, if the candidate is a stationary background object, the expected variation of offset positions is shown (solid line). This is based on a distance and on a primary proper motion, as well as the initial offset position of the candidate relative to the primary. The predicted offset positions of a stationary background object for the second epoch is shown in red with uncertainties. For HIP117452, measurements of both components Ba and Bb at various epochs are plotted in dark and light blue, respectively.
}
\label{astrometry}
\end{figure*}

Follow-up observations of the candidates were automatically scheduled and obtained using the DBI mode with $J_2J_3$ filters of IRDIS, which is well adapted to distinguish background stars from physically young, early-T, or warm mid-L dwarf planets, and offers an additional epoch for a proper motion test. Follow-up observations were then processed using SpeCal with the TLOCI algorithm in ADI only and a template approach as before. All companion candidates were re-detected, except the one around HIP\,8832 falling outside the IRDIS field, given its large separation and an observing sequence that was not perfectly centered with the meridian passage. The results are reported in Table\,\ref{table_candidates}. The use of a different pair of filters enabled us to explore the companion candidate photometric properties in the $J_2$-band- and $J_3$-band-based CMD, for which we also report the distribution of background stars observed in previous crowded fields (see Figure\,\ref{diag_mag}, \textit{Right}). One can directly see that most of our late-L to early-T potential companion candidates, including the previous ones identified as stationary background stars around HIP\,34276 (cc1 and cc2) and HIP\,101800 (cc1 and cc2), fall onto the background contaminant sequence indicating that they are most likely background stars. As a further check, we used the relative astrometry obtained at two epochs to estimate the proper motion of the companion candidates relative to their primary stars. Figure\,\ref{astrometry} shows the proper motion plots of each candidate and confirms that the companion candidates around HIP\,34276, HIP\,101800, and HIP\,95619 are not co-moving with their primary stars. The distance and proper motion of the stars, with their uncertainties, are taken from the \textit{Gaia} Data Release 2 catalog \citep{Gaia2018}. For HIP\,16095, given the relatively low proper motion of the star, the status of the companion candidate HIP\,16095-cc1 remains ambiguous. However, the $J_2$-band- and $J_3$-band-based CMD still supports a background contamination. If bound, this candidate would have an estimated mass between 7 and 12 $M_{Jup}$ at the system age ($\leq100$ Myr) and distance (88\,pc) illustrative of the SPHERE detection performances around young nearby stars beyond 10\,au.

\begin{figure*}[t]
\centering
\includegraphics[width=0.435\textwidth]{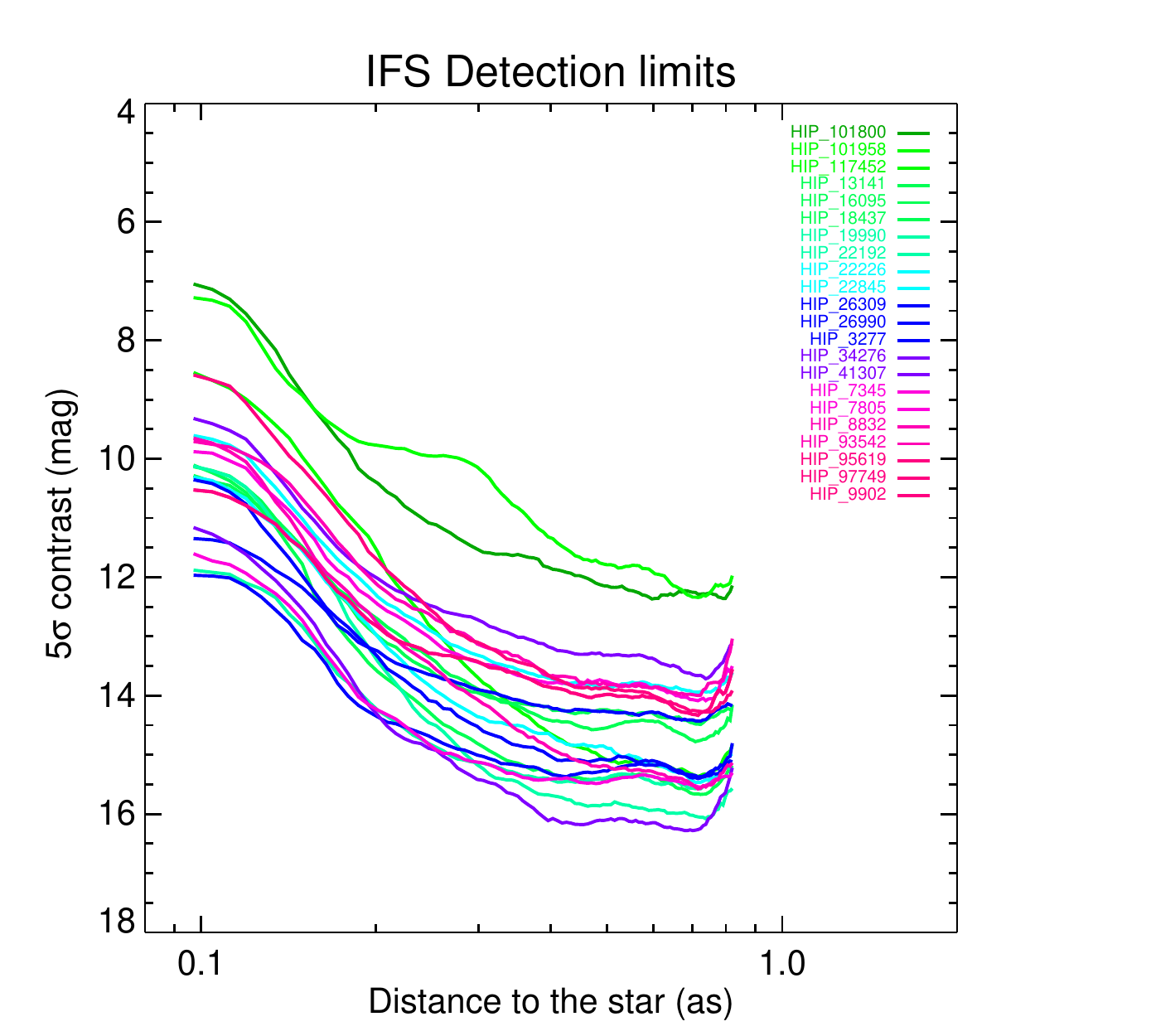}
\hspace{0.2cm}
\includegraphics[width=0.435\textwidth]{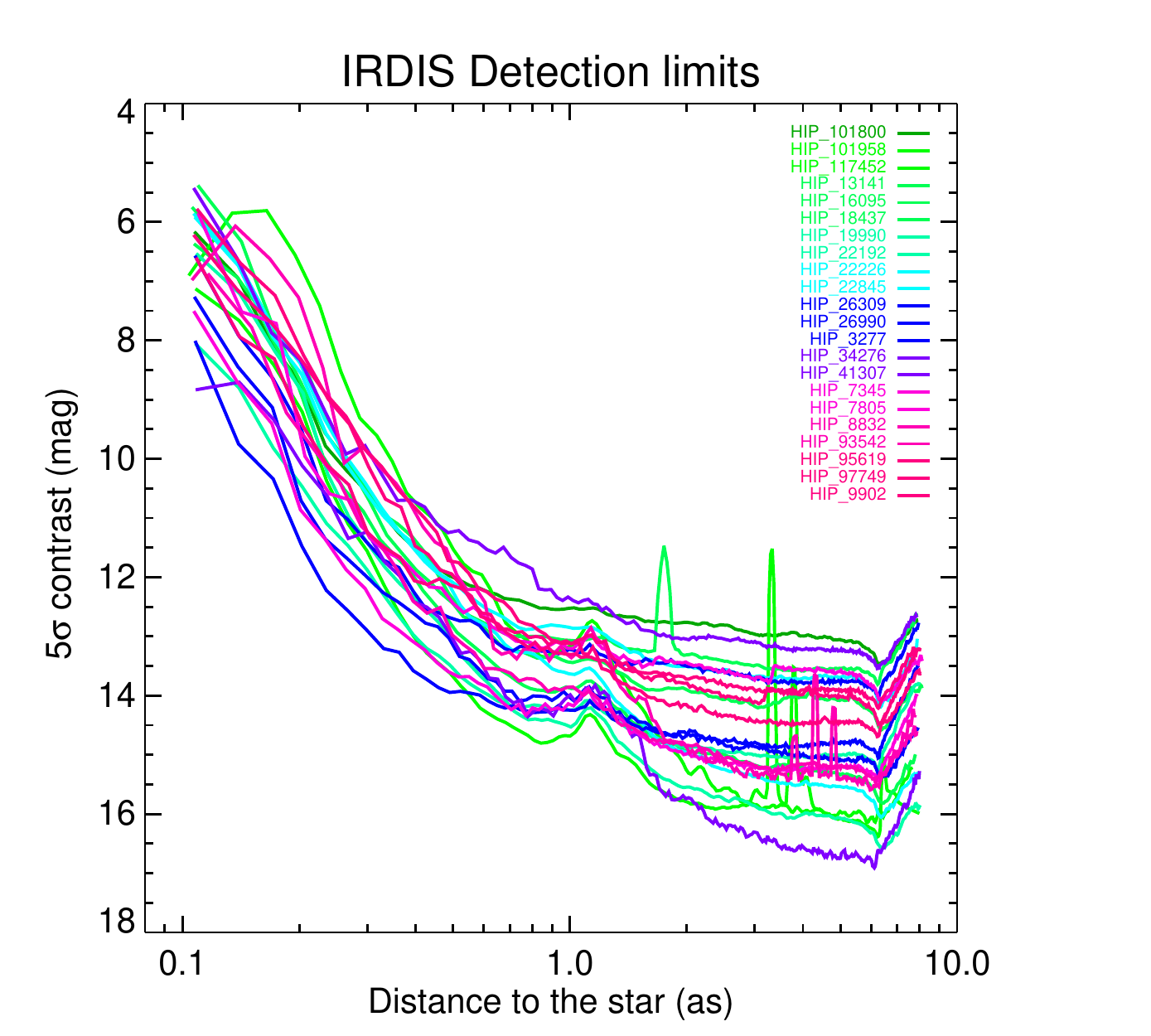}
\caption{Magnitude contrast limit curves for all targets with TLOCI algorithm.}
\label{detlim}
\end{figure*}

For HIP\,117452 Ba and Bb, the colors and magnitudes in $K_1$ and $K_2$ compared to the predictions of the evolutionary models of \cite{Siess2000} suggest that Ba and Bb are a pair of M1 and M2 low-mass stars, considering an age of 40\,Myr at a distance of 42\,pc. Combining our relative astrometry with the one reported by \cite{Matthews2018} and shown in Table\,~\ref{table_candidates}, we performed a first orbit fitting of the pair. Following the method developed by \cite{Chauvin2012}, we used a Markov chain Monte Carlo (MCMC) Bayesian analysis technique \citep{Ford2007}, which is well suited for observations covering a small part of the whole orbit (for large orbital periods). We did not consider any prior information using the proximity of the primary star. The results are reported in Figure\,~\ref{mcmc} and favor a relatively inclined orbit $i\sim98_{-5}^{+8}$\,deg, a longitude of ascending node fairly well-constrained at $\Omega=20\pm2$\,deg, tight semi-major axis $a\sim14_{-4}^{+7}$\,au, but surprisingly large eccentricities $e\ge0.4$. These large values of eccentricity are not dynamically expected, given the proximity of the primary star located at a physical projected separation of $\sim150$\,au, although the orbit of the binary companion around HIP\,117452 is not known. Fitting solutions using a least squares Levenberg-Marquardt (LSLM) algorithm \citep{Press1992} to search for the model with the minimal reduced $chi^2$ are also reported for comparison. Further dynamical study of the global system considering the debris disk architecture around HIP\,117452 and the binary companion HIP\,117452\,BaBb configuration is be needed.

\section{Detection limits and survey completeness}
\label{sec:detection_limits}
To exploit the information from the actual nondetection in IFS and IRDIS observations of the survey, the detection limits of each individual observations were then estimated. Based on SpeCal results, we derived a standard pixel-to-pixel noise map for each observing sequence corrected from the flux loss related to the ADI or ASDI processing by injecting fake planets. The detection limit maps at $5\sigma$ were then obtained using the pixel-to-pixel noise map divided by the flux loss and normalized by the relative calibration with the primary star (considering the different exposure times, the neutral density, and the coronograph transmission). These detection limits were finally corrected from small number statistics following the prescription of \cite{Mawet2014} to adapt our $5\sigma$ confidence level at small angles with IRDIS and IFS. The $5\sigma$ contrast curves, resulting from the azimuthal average of the detection maps, are reported for IFS and IRDIS in Figure~\ref{detlim}.

\begin{figure}[t]
\centering
\includegraphics[width=\columnwidth]{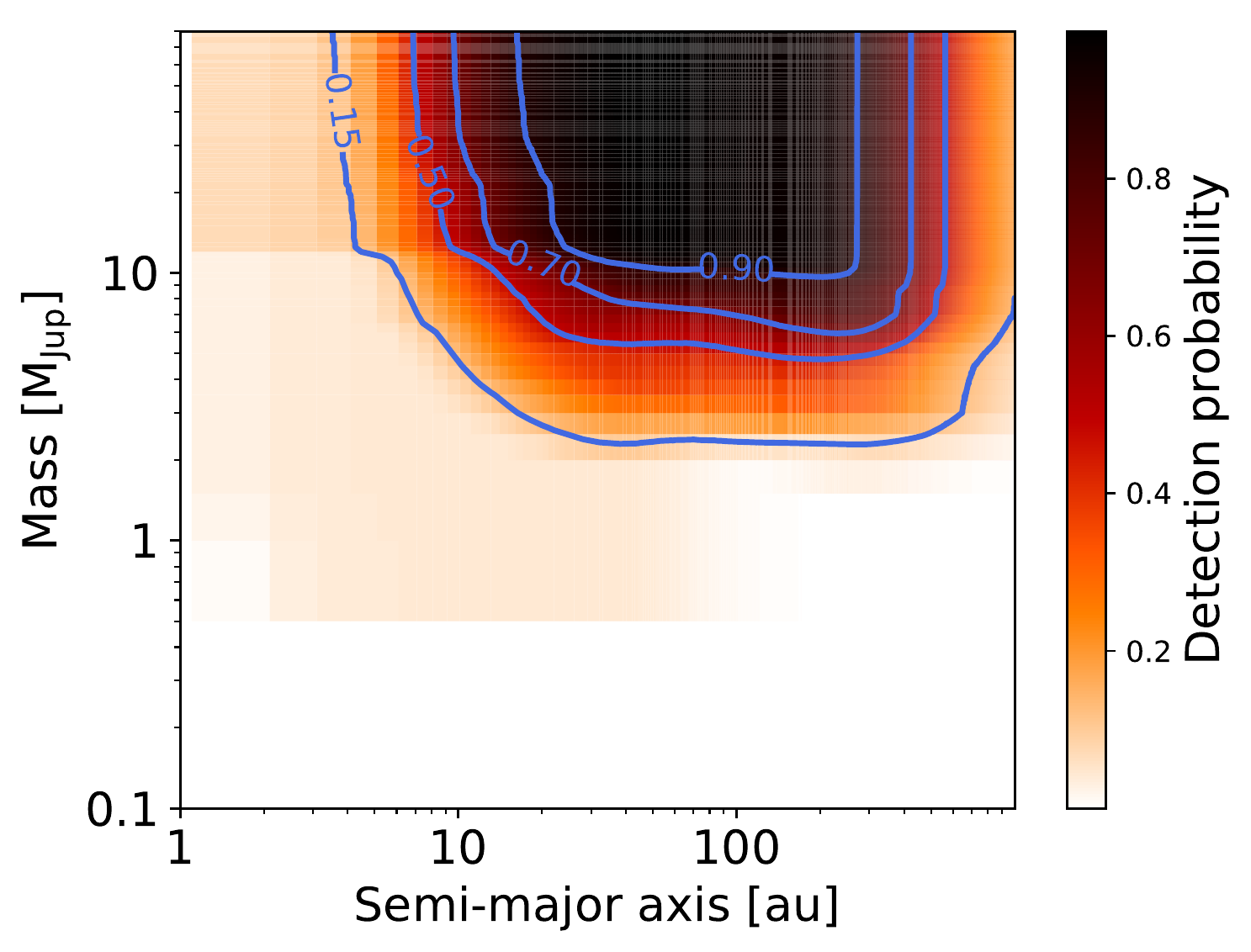}
\caption{Combined mean detection probability map for the whole survey.}
\label{mass_limit_mean}
\end{figure}

To convert the detection limits in terms of the mass and semi-major axis parameter space explored with SPHERE, we used the multi-purpose exoplanet simulation system (MESS) code, a Monte Carlo tool for the statistical analysis and prediction of exoplanet search results \citep{Bonavita2012}. This code has been used extensively in previous direct imaging surveys for that same purpose \citep{Chauvin2010,Chauvin2015,Chauvin2018,Vigan2012,Vigan2017,Rameau2013,Lannier2016}. With MESS, we then generated a uniform grid of mass and semi-major axis in the interval [1, 80]~M$_{\rm{Jup}}$ and [1, 1000]~au with a sampling of 0.5~M$_{\rm{Jup}}$ and 1~au, respectively. \\

For each point in the grids, 100 orbits were generated, randomly oriented in space from uniform distributions in cos(i), $\omega$, $\Omega$, $e \le 0.8$, and $T_p$. We built detection probability maps by counting the number of detected planets over the number of generated ones and simply comparing the on-sky projected position (separation and position angle) of each synthetic planet with the SPHERE 2D detection limit maps at $5\sigma$ converted in masses based on the COND (hot-start) model predictions \citep{Baraffe2003}. The primary age, distance, and magnitude reported in Table\,\ref{table_p99} are considered for the luminosity-mass conversion.\\

The resulting detection probability map of the complete survey is reported in Figure\,\ref{mass_limit_mean}. This result shows that, despite the relatively wide age range (20 to 120\,Myr) and distance (10 to 102\,pc) of our sample, we achieved a relatively good detection probability larger than $50\,\%$ for giant planets with masses larger than 5\,~M$_{\rm{Jup}}$ and semi-major axes between 10 and 500\,au, sufficient for the detection of system analogs to HR\,8799 or HD\,95086. In principle, the degeneracy between mass and initial entropy could change these limits considerably (e.g., \citealp{Marleau2014,Brandt2014}). In practice, however, taking more realistic post-formation entropies into account strongly mitigates this problem, as shown for instance in the case of HIP~65426~b by \citet{Marleau2019}.

\begin{figure*}[t]
\centering
\includegraphics[width=\textwidth]{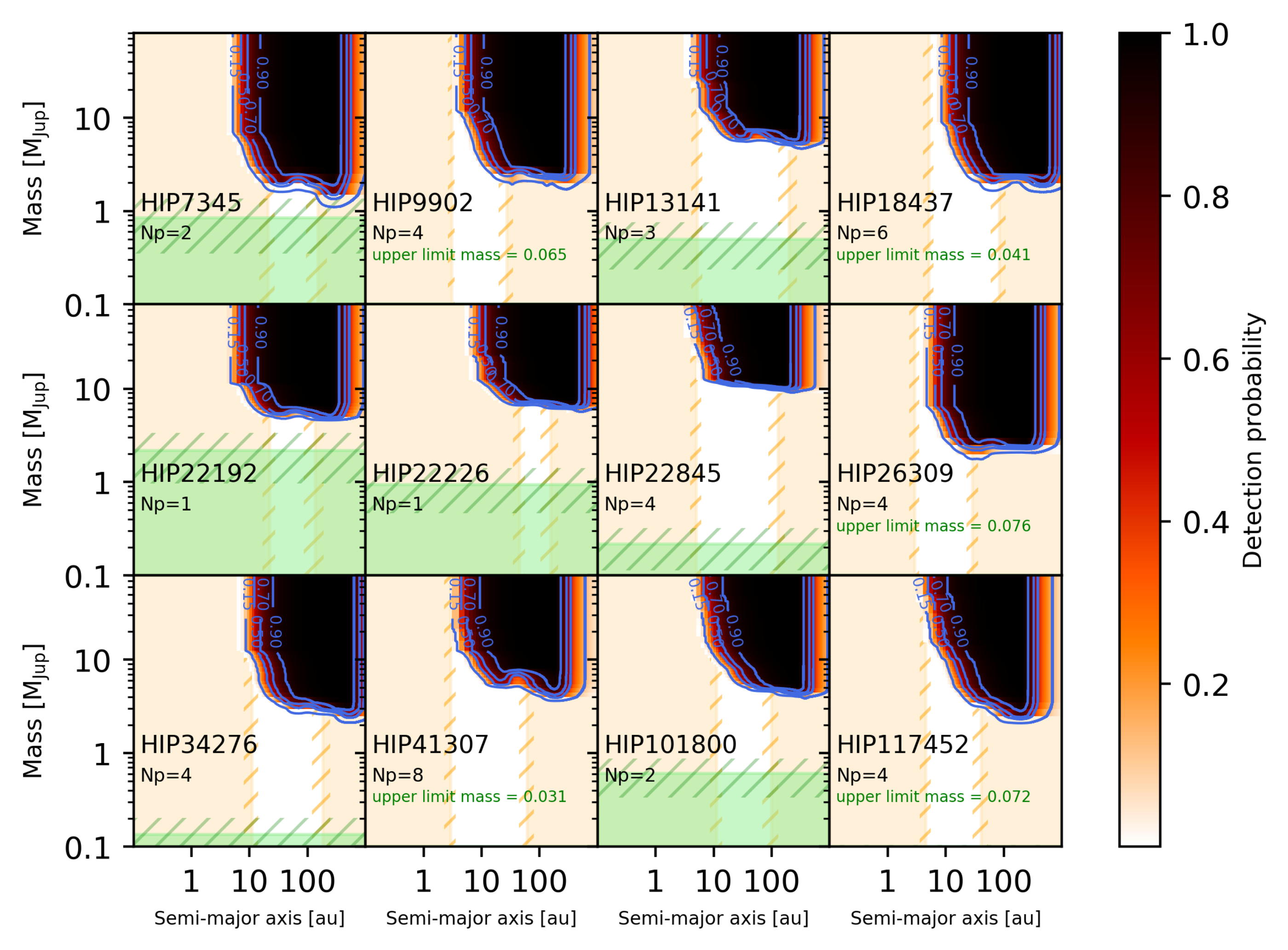}
\caption{Constraints on planetary systems for 12 targets in our survey. The positions of the inner and the outer debris belts are shown in orange by shading the regions inside the inner and beyond the outer. Our mass contrast limits are based on SPHERE/IRDIS and COND model predictions. Dynamical mass constraints for a slightly closer planet spacing of 20 mutual Hill radii from \cite{Shannon2016} are shown in green with masses below this value shaded. Np is the number of planets with the mass (indicated in green) required to open the gap. The uncertainties for debris belt position and dynamical mass limit are calculated based on the uncertainty in debris belt temperature, and indicated with hatching. }
\label{mass_limit}
\end{figure*}

\section{Discussion}
Our survey is composed of relatively old  gas-free systems. Therefore, some of these systems contain debris disks. We assumed that planets are a valid explanation for the formation of debris structure, as shown in the case of the Solar System where planets are known to reside between two belts of debris, and in the case of HR\,8799 and HD\,95086 where planets are known to reside in two-temperature debris disks. The analysis of our survey follows the work by \cite{Matthews2018}. The temperature values of debris belts are found in \cite{Chen2014}, where these temperatures were estimated using a two-temperature black-body model and a Bayesian parameter estimation to select the best model to fit the SED. The disk radii were calculated following \cite{Pawellek2015}, assuming that dust are composed of 50\% astrosilicate and 50\% ice. In addition, we constrained our SPHERE/IRDIS observations with dynamical arguments on the possible planetary systems hiding within the debris gaps \citep{Shannon2016}.\\

Mass limits were calculated with the MESS code as described in Section~\ref{sec:detection_limits} and shown in Figure\,\ref{mass_limit}. The theoretical mass for a single planet to clear the observed gap is large $\geq 25 M_\mathrm{J}$ \citep{Nesvold2014}. Therefore, in our cases, we infer that the systems must be in multi-planet configuration, as in HR8799, in which several planets with lower masses clear the gap. In Figure~\ref{mass_limit}, we plot the minimum masses of planets required to clear the debris gaps, as well as their location and their "Np" number based on the N-body simulations of \cite{Shannon2016}. This model considers only planets with low eccentricities. The mass and the Np number change if the eccentricity is larger. The mass, shown in green in Figure~\ref{mass_limit}, is the minimum mass per planet, with uncertainties based on the age of the system and on the belts radius. The minimum mass calculation assumes that planets are spaced by a typical value of $\sim 20$ mutual Hill radius ($R_\mathrm{H}$), which is consistent with the value of $21.7 \pm 9.5 \mathrm{R_H}$ predicted by \cite{Fang2013}. \\

By combining the observational upper and theoretical lower mass constrains, only a small region of parameter space is unconstrained. For 12 targets in our survey of which the temperature values are found in \cite{Chen2014}, we infer a multi-planet system based on the large theoretical clearing masses. In such a multi-planet system, the widest separation planet would have a physical separation close to that of the outer debris belt, where our direct imaging limits are relatively tight. In main cases, planets must be at least $\sim 0.1M_J$ to clear the observed gap based on dynamical arguments, and in some cases the dynamical mass limit exceeds $1M_J$. In Figure~\ref{mass_limit}, for the target HIP7345, the mass limit, $\sim 1.3 M_J$ at 90\% in the gap, is close to the dynamical mass limit $\sim 0.9$. \\

Among our 12 targets for which we note the presence of two debris belts, no exoplanetary mass companions were detected. Our sample is too small for a detailed statistical analysis. However, a nondetection in a sample of 12 stars is not inconsistent with the debris disk occurrence rate of 6.27\% in a debris disk sample of planets between $5-20 M_{\mathrm{J}}$ and 10-1000\,au \citep{Meshkat2017}, since we would expect that some companions might be below our detection limits. Our nondetections are also consistent with the lower occurrence rate of $\sim 1 \%$ found in \cite{Bowler2016} and \cite{Galicher2016}. The results of this 12 target sample are not incompatible with the theory that planets are carving wide debris gaps, since in each case our direct imaging mass limits are higher than the theoretical mass limits that we calculate.\\

The existence of the planetary perturbers beyond 5\,au, and potentially these architectures will be explored in futur observations: i/ observations combining radial velocity, astrometry with \textit{GAIA} for the inner parts ($\le5$\,au), ii/ observations with the next generation of planet imagers from the ground (SCExAO, KPIC, SPHERE+, GPI2.0 on 10m-class Telescopes, then with the ELTs) and space (\textit{JWST}, \textit{WFIRST}).

\section{Conclusions}
We reported the observations and analysis of a survey of 22 stars with VLT/SPHERE with IRDIS in the DBI mode with $K_1K_2$ filters and  $J_2J_3$ for the follow-up observations, and IFS in the $Y-H$ filters, with the goal of detecting and characterizing giant planets on wide orbits. The selected sample favors young, that is to say$\leq 100$ Myr, nearby, $\leq 100$ pc, dusty, and early-type stars to maximize the range of mass and separation, over which the observations are sensitive. The optimized observation strategy with the angular differential imaging in thermal bands and a dedicated data reduction using various algorithms allow us to reach a typical contrast  12.5 mag at 0.25” and 14 mag at 1.0” in IRDIS. These contrasts are converted to mass limits for each target.  Despite the good sensitivity of our survey, we did not detect any new giant planets. We confirmed that the sources detected around HIP\,34276, HIP\,101800, HIP\,16095, and HIP\,95619 are stationary background sources by analyzing $K_1$-band, $K_2$-band, $J_2$-band, and $J_3$-band images and their relative motions. The status of the candidate around HIP\,8832 still requires further follow-up. HIP\,117452\,BaBb is resolved and confirmed as a binary companion \citep{Derosa2011,Matthews2018}. For 12 targets of our survey, where we determined the radii of the debris belt, we derived upper and lower mass limits. We used Monte Carlo simulations to estimate the sensitivity survey performance in terms of planetary mass and semi-major axis to perform the upper limit. We additionally calculated the minimum required mass for planets in the system to have cleared the observed debris gap to perform the lower mass limit. Combining our upper and lower mass limits, we are able to tightly constrain the unexplored parameter space around these systems: typically, planets must be at least $\sim 0.1M_J$ in main cases to clear the observed gap based on dynamical arguments, and in some cases the dynamical limit exceeds $1M_J$. Direct imaging data from VLT/SPHERE are sensitive to planets of $\sim 3 M_J$ for a typical target in our survey. Several of the planetary systems will likely be detectable with the next generation of high-contrast imagers.

\begin{acknowledgements}
First, we thanks the referee for providing useful comments.
This project was partly supported by the IDEXLyon project (contract ANR-16-IDEX-0005) under the auspices University of Lyon. It was supported by CNRS, by the Agence Nationale de la Recherche (ANR-14-CE33-0018). It has received funding from the European Union’s Horizon 2020 research and innovation programme under the Marie
Skłodowska-Curie grant agreement No 823823. G-D. Marleau acknowledges the support of the DFG priority program SPP 1992 ``Exploring the Diversity of Extrasolar Planets" (KU 2849/7-1). C.~Mordasini and G.-D.~Marleau acknowledge support from the Swiss National Science Foundation under grant BSSGI0\_155816 ``PlanetsInTime''. Parts of this work have been carried out within the frame of the National Centre for Competence in Research PlanetS supported by the SNSF. A. Bayo acknowledges support from ICM (Iniciativa Cient\'ifica Milenio) via the N\'ucleo Milenio de Formaci\'on Planetaria, and from FONDECYT (grant 1190748). Finally, this work has made use of the the SPHERE Data Centre, jointly operated by OSUG/IPAG (Grenoble), PYTHEAS/LAM/CESAM (Marseille), OCA/Lagrange (Nice), Observatoire de Paris/LESIA (Paris), and Observatoire de Lyon, also supported by a grant from Labex  OSUG@2020 (Investissements d’avenir – ANR10 LABX56). 

\end{acknowledgements}


\bibliographystyle{aa}
\bibliography{biblio}

\onecolumn
\begin{appendix} 
\section{MCMC orbital fit of HIP\,117452\,BaBb}
\begin{figure*}[h] 
\centering
\includegraphics[width=16cm]{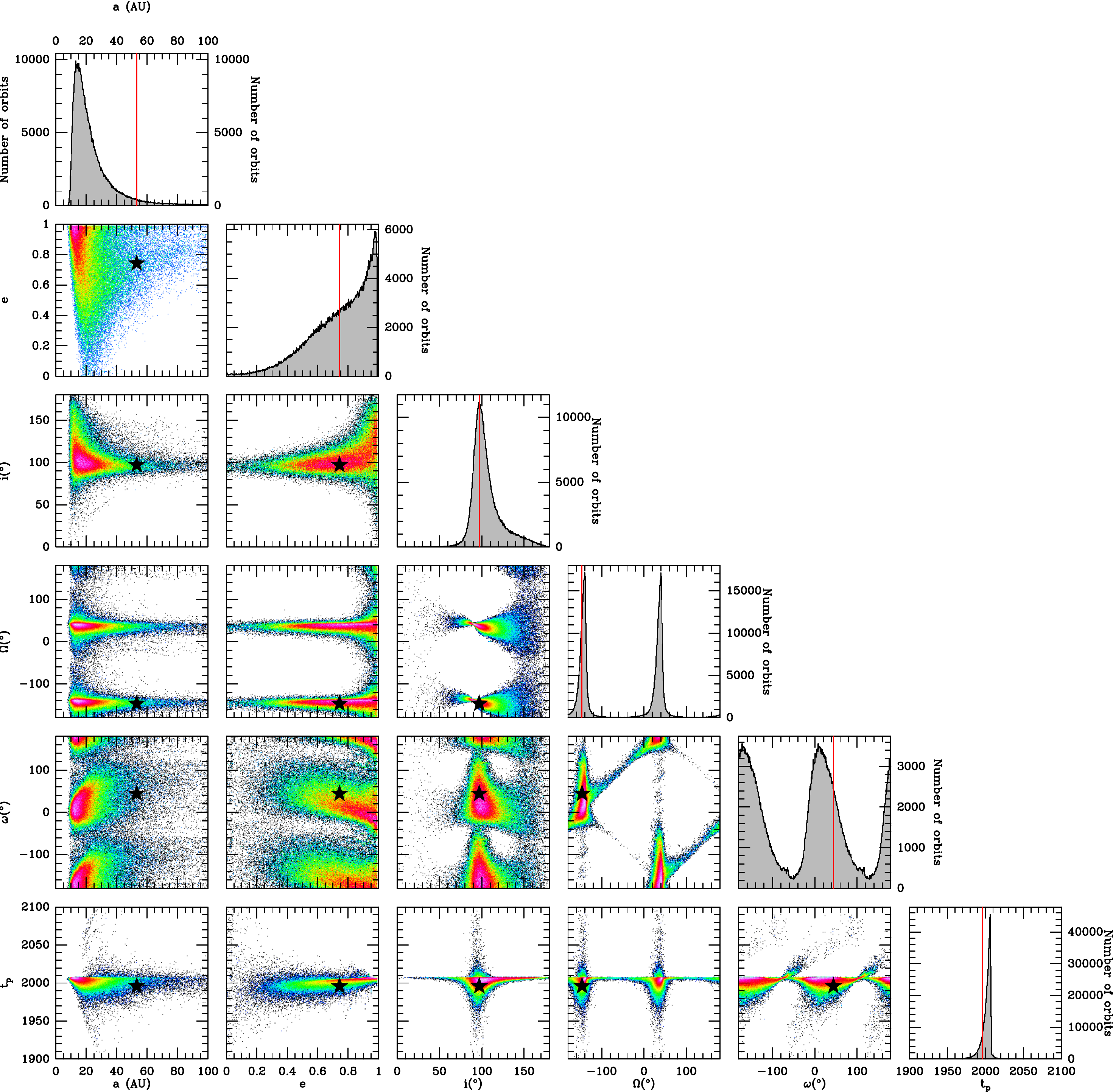}
\caption{Results of the MCMC fit of the NaCo and SPHERE combined astrometric data of HIP\,117452 Ba and Bb reported in terms of statistical distribution matrix of the orbital elements $a$, $e$, $i$,$\Omega$,$\omega$, and $t_p$. The red line in the histograms and the black star in the correlation plots indicate the position of the best LSLM-$\chi^2$ model obtained for comparison.}
\label{mcmc}
\end{figure*}

\end{appendix}

\end{document}